\documentclass[aps,pre,twocolumn,groupedaddress]{revtex4-1}

\usepackage{graphicx}

\begin{document}


\title{Nonequilibrium Green's function method for quantum thermal transport}


\author{Jian-Sheng Wang}
\email[]{phywjs@nus.edu.sg}
\homepage[]{http://staff.science.nus.edu.sg/~phywjs/}
\author{Bijay Kumar Agarwalla}
\author{Huanan Li}
\author{Juzar Thingna}
\affiliation{Department of Physics and Centre for Computational Science and Engineering, National University of Singapore, Singapore 117542, Republic of Singapore}


\date{29 March 2013}

\begin{abstract}
This review deals with the nonequilibrium Green's function (NEGF) method applied to the problems of energy transport due to atomic vibrations (phonons), primarily for small junction systems.  We present a pedagogical introduction to the subject, deriving some of the well-known results such as the Laudauer-like formula for heat current in ballistic systems. The main aim of the review is to build the machinery of the method so that it can be applied to other situations, which are not directly treated here.  In addition to the above, we consider a number of applications of NEGF, not in routine model system calculations, but in a few new aspects showing the power and usefulness of the formalism.  In particular, we discuss the problems of multiple leads, coupled left-right-lead system, and system without a center. We also apply the method to the problem of full counting statistics.  In the case of nonlinear systems, we make general comments on the thermal expansion effect, phonon relaxation time, and a certain class of mean-field approximations.  Lastly, we examine the relationship between NEGF, reduced density matrix, and master equation approaches to thermal transport.  
\end{abstract}


\maketitle

\section{Introduction}
The method of nonequilibrium Green's functions (NEGF) was initiated by Schwinger 
in a rather mathematical paper \cite{schwinger61} for a treatment of Brownian motion of a quantum oscillator.  Already in 1961, the importance of forward and backward evolution in the calculation of nonequilibrium quantum expectation values at time $t$ evolved from an earlier time was recognized and the six different Green's functions defined.  The next important development in NEGF came due to 
Kadanoff and Baym \cite{kadanoff62} where the main emphasis was to derive quantum kinetic equations.  Keldysh showed that diagrammatic expansion is possible even for nonequilibrium processes \cite{keldysh65}, a key idea being contour order.  These initial developments all occurred in the early 1960s.  There are a number of earlier reviews \cite{chou84,danielewicz84,rammer-RMP-86} and conference series  \cite{bonitz00,bonitz03} that people working in this field should be aware of. An important paper on treating transport by NEGF is that of Caroli, et al. \cite{caroli71}, where for the first time, an explicit formula for the transmission coefficient  in terms of the Green's functions was given.  Its modern form presented here is due to Meir and Wingreen \cite{meir92}.  Some of the very recent reviews on NEGF method, mostly still for electronic transport, can be found in Refs.~\onlinecite{prociuk08,aeberhard11,zimbovskaya11,nikolic12}.  

This paper can be thought as an update to our earlier review \cite{wang08review} on the application of NEGF to phonon transport. The main aim is to develop the theory more systematically and to review the various new applications.  Some of the straightforward, routine recent applications, e.g., Refs.~\onlinecite{lan09,hopkins09,xie11,tian12,bachmann12,yeo12}, will not be discussed.  We start slow with the problem of the harmonic oscillator in Sec.~II.  Since any phononic systems at the ballistic level can be thought of as coupled oscillators, and in eigenmodes, independent oscillators, the single mode oscillator is fundamental to the NEGF method.  In Sec.~III, we define the nonequilibrium Green's function proper.  Here, we look at the contour ordered Green's function as well as operators used to define it by carefully introducing a contour version of the evolution operator as well as giving a formal definition of the Heisenberg operator on the contour.
The mathematical aspect of functions defined on the contour is dealt with in Sec.~IV.  Two methods are available for obtaining equations and actual computation, the equation of motion method and Feynman diagrammatic expansion.  Both of them are formulated on the contour of a finite segment $[t_0, t_M]$.  This is discussed in the following two sections, V and VI.  We emphasize the view that the contours are defined on finite segment.  This point of view makes the theory valid both for transient and steady state.   The current formulas are derived in Sec.~VII.   The remaining sections review some applications, including problems of multiple leads, full counting statistics, which is to look at the full distributions of transferred energy in a given time interval.  We review few applications in nonlinear situations where NEGF gives reasonably good results, this includes thermal expansion and phonon life time, and a self-consistent mean-field theory for a quartic nonlinear junction.  NEGF is normally concerned with Green's functions, but it can also say much on the reduced density matrix; here in Sec.~XII, we review Dhar, Saito, and H\"anggi's method of computing the reduced density matrix in steady state for a transport system.  This is quite relevant with respect to the last topic of this review, the quantum master equation approach.  We try to rephrase the usual quantum master equation in terms of NEGF and offer an approach and formula to obtain higher order current with respect to the system-bath couplings. 
We end the review with a brief summary in Sec.~XIV.

\section{Harmonic Oscillator, Equilibrium Green's Functions}
In this review, we take a bottom up approach to `build' the nonequilibrium Green's functions from the equilibrium ones.  This will be done in the first few sections. In this section, we review the basic properties of a single degree of freedom harmonic oscillator in thermal equilibrium.  The set of functions defined here will be found of great utility later as any phononic system (and even photonic systems) can be thought of as a collection of independent harmonic oscillators if we work in the eigenmodes.   Hence the problem of the harmonic oscillator is fundamental to phonon transport.  Similar discussions and formulae can also be found in the lecture notes of Brouwer \cite{brouwer05}, Sec.~2.6 (an excellent introduction to NEGF),  in Kleinert  \cite{kleinert09}, Chap.~18.5.1 (although the conventions are different from ours), as well as in appendix~B of Ref.~\onlinecite{jiang-prb80-205429} and in appendix~A of Ref.~\onlinecite{argarwalla-pre12}.  

We assume that the reader is familiar with the solution of the quantum-mechanical problem of a harmonic oscillator using creation/annihilation operators (see, e.g.,  B\"ohm \cite{bohm79}, Chap.~II.3).  The Hamiltonian of a single quantum oscillator is given by
\begin{equation}
H = \frac{p^2}{2m} + \frac{1}{2} k x^2,
\end{equation}  
where $x$ is the displacement operator, $p$ is the conjugate momentum such that
$[x,p]=xp-px=i \hbar$, $\hbar$ is the reduced Planck constant, $m$ is mass, and $k$ is the force constant.  For notational simplicity, in this review, we'll always perform the transform $u = x \sqrt{m}$ so that the mass can be transformed away. Also introducing $k = m \Omega^2$, where $\Omega>0$ is the oscillator angular frequency, the Hamiltonian can be rewritten as
\begin{equation}
H =\frac{1}{2} \dot{u}^2 + \frac{1}{2} \Omega^2 u^2 = 
\hbar \Omega\left( a^\dagger a + \frac{1}{2} \right),
\end{equation} 
where the mass-normalized displacement $u$ can be expressed in terms of the annihilation and creation operators as
\begin{equation}
\label{eq-u}
u = \sqrt{\frac{\hbar}{2 \Omega}}\; \bigl(a + a^\dagger\bigr).
\end{equation}
We have the commutation relation $[a, a^\dagger]=1$. In the Heisenberg picture, the operators evolve in time, and the states do not change.  The Heisenberg equation of motion for $a$ takes a very simple form
\begin{equation}
\label{eq-dadt}
{d a(t) \over dt} = \frac{1}{i\hbar} [a(t), H] = -i \Omega\, a(t).
\end{equation}
This gives the oscillatory solution $a(t) = a\, e^{-i \Omega t}$, thus the Heisenberg
solution for $u$ can be easily obtained.

In equilibrium statistical mechanics, we assume that the system is not in a pure quantum state, but in various states with some probabilities.  More precisely, we need to describe the system with a density matrix (see Huang \cite{khuang87}, Chap.~8).  In this review, we'll always use the canonical density operator, $\rho = 
e^{-\beta H}/{\rm Tr}(e^{-\beta H})$, $\beta = 1/(k_B T)$, where $k_B$ is the Boltzmann constant and $T$ is the absolute temperature.  In the energy eigen-state representation, $|n \rangle$, 
the Hamiltonian is diagonal,  and using the facts
\begin{eqnarray}
H |n \rangle &=& \bigl(n + \frac{1}{2}\bigr)  \hbar \Omega|n \rangle, \\
a |n \rangle &=& \sqrt{n} | n\!-\!1 \rangle,\\
a^\dagger |n \rangle & = & \sqrt{n+1} |n \!+\!1 \rangle,
\end{eqnarray}
we find
\begin{eqnarray}
\label{eq-aa}
\langle a a \rangle = 0, \quad \langle a^\dagger a^\dagger \rangle = 0, \\
\label{eq-adagger}
\langle a^\dagger a \rangle = f, \quad \langle a a^\dagger  \rangle = 1+f,
\end{eqnarray}
where the angular brackets denote trace with the canonical density matrix,
i.e., $\langle \cdots \rangle = {\rm Tr}(\rho \cdots)$, and 
\begin{equation}
f = {1 \over e^{\beta\hbar \Omega} - 1}
\end{equation}
is the Bose-Einstein distribution function.

Now we are ready to define correlation functions, or Green's functions for the harmonic oscillator. One can define the Green's functions using the creation/annihilation operators --- this is traditionally done in many-body theory.  But for phononic systems, it is more efficient if we only use the displacement operators.  We define the greater Green's function as
\begin{equation}
g^{>}(t,t') = - \frac{i}{\hbar} \langle u(t) u(t') \rangle,
\end{equation}
where the time-dependence is from the Heisenberg evolution, and the angular bracket is for the average over the equilibrium density matrix $\rho$.  Using the relation between $u$ and $a$, Eq.~(\ref{eq-u}), and the solution of Eq.~(\ref{eq-dadt}), and Eq.~(\ref{eq-aa}), (\ref{eq-adagger}), we get
\begin{equation}
\label{eq-ggreater}
g^{>}(t,t') = - \frac{i}{2\Omega} \left[ f e^{i\Omega(t-t')} + (1+f) e^{-i\Omega(t-t')} \right].
\end{equation}
We note that the function $g^{>}(t,t') = g^{>}(t-t')$ is actually a function of one argument due to time translational invariance.  It is always so provided that the system is in thermal equilibrium or in a nonequilibrium steady state.   The greater Green's function is nothing but the position-position correlation function in time.  The extra factor $(-i/\hbar)$ is there to match the evolution operator in a Dyson expansion and is purely conventional.  With this definition, $g^{>}(t)$ has the dimension of time. We are going to define a few more functions: the lesser Green's function
\begin{equation}
g^{<}(t,t') = - \frac{i}{\hbar} \langle u(t') u(t) \rangle = g^{>}(t',t),
\end{equation}
the time-ordered (causal) Green's function
\begin{eqnarray}
g^{t}(t,t') &=& - \frac{i}{\hbar} \langle T u(t) u(t') \rangle \nonumber\\ 
&=& \theta(t-t') g^{>}(t,t') + 
\theta(t'-t) g^{<}(t,t'),
\end{eqnarray}
and the anti-time-ordered Green's function
\begin{eqnarray}
g^{\bar{t}}(t,t') &=& - \frac{i}{\hbar} \langle \bar{T} u(t) u(t') \rangle \nonumber\\ 
&=& \theta(t'-t) g^{>}(t,t') + 
\theta(t-t') g^{<}(t,t').
\end{eqnarray}
The Heaviside step function is defined as $\theta(t) = 1$ if $t\ge 0$ and 0 otherwise.
The meaning of the time-order operator $T$ and anti-time-order operator $\bar T$
is given by the second line.  For the time order, the positions of the two operators are unspecified until the time $t$ and $t'$ are known.  The position of the operators is such that the operator on the right is the earliest and following the order of time as one goes from right to left.  The anti-time order is the opposite.

It seems redundant at this point to introduce four of these functions, as Eq.~(\ref{eq-ggreater}) determines all the others.  However, these four functions form the components of so-called contour ordered Green's function, $g(\tau, \tau')$, which has great utility when we deal with nonequilibrium situations.   Another pair of
important Green's functions are the retarded Green's function, given by
\begin{eqnarray}
g^{r}(t,t') &=& - \frac{i}{\hbar}  \theta(t-t') \bigl\langle [u(t), u(t')] \bigr\rangle \nonumber\\
&=& -\theta(t-t') { \sin \Omega(t-t') \over \Omega }, 
\end{eqnarray}
and the advanced Green's function
\begin{equation}
g^{a}(t,t') = \frac{i}{\hbar} \theta(t'-t) \bigl\langle [u(t), u(t')] \bigr\rangle.
\end{equation}
The retarded Green's function appears in linear response theory, and it has the same meaning as that of Green's function in classical physics, i.e., it is the solution of the
equation
\begin{equation}
\ddot{g}^r(t) + \Omega^2 g^r(t) = - \delta(t),
\end{equation}
with the condition $g^r(t) = 0$ for $t<0$, where $\delta(t)$ is the Dirac $\delta$ function.

In practical calculation, particularly in the case of time-translationally invariant situation, it is more convenient to work in the frequency domain.  We thus define the Fourier
transform of the Green's functions, e.g., by 
\begin{equation}
g^r[\omega] = \int_{-\infty}^{+\infty}\!\!\! g^r(t) e^{i\omega t} dt,
\end{equation}
and its inverse 
\begin{equation}
g^r(t) = \int_{-\infty}^{+\infty}\!\!\! g^r[\omega] e^{-i\omega t} {d\omega \over 2 \pi}.
\end{equation}
We use the same symbol for a function of time defined in the whole domain
$(-\infty, +\infty)$ and its Fourier transform.  Whether it is the function of time
or its Fourier transform is indicated by its argument $(t)$ or $[\omega]$.  Just like the time $t$, since $\omega$ is a Fourier transform variable, it also varies in the
domain from $-\infty$ to $+\infty$.  The Fourier transform of the retarded Green's function for a single oscillator is
\begin{eqnarray}
g^r[\omega] &=& -\int_{-\infty}^{+\infty}\!\!\!\! \theta(t) {\sin (\Omega t) \over \Omega} e^{i\omega t  - \eta t} dt \nonumber \\
&=& { 1 \over (\omega + i\eta)^2 - \Omega^2 }, \quad (\eta \to 0^+).
\end{eqnarray}
It is important to add a small positive damping factor $\eta$ so that the integral converges.  This choice displaces the poles in the complex plane of frequency below the real axis and produces the desired causality property that
$g(t) = 0$ for $t<0$ when one performs an inverse Fourier transform using contour integral.  The advanced version is obtained by complex conjugate, $g^a[\omega] = g^r[\omega]^{*}$. 

The lesser Green's function in frequency domain is
\begin{equation}
g^{<}[\omega] = -\frac{i\pi}{\Omega} \left[
f \delta(\omega - \Omega) + (1+f) \delta(\omega + \Omega) \right].
\end{equation}
Using the Plemelj formula
\begin{equation}
{ 1 \over x + i \eta } = P \frac{1}{x} - i \pi \delta(x),
\end{equation}
where $P$ stands for Cauchy principal value, we can now relate the lesser Green's function with the retarded Green's function as
\begin{equation}
\label{eq-fluctuation-dissipation}
g^{<}[\omega] = \bigl(g^r[\omega] - g^a[\omega]\bigr) f(\omega).
\end{equation}
This equation turns out to be true in general (in the sense of the $G$ defined in the next section) for equilibrium systems and is one particular form of a fluctuation-dissipation relation.
Another, which is actually equivalent to that one, is 
$g^>[\omega] = e^{\beta \hbar\omega} g^<[\omega]$.  In the time domain, this is the 
Kubo-Martin-Schwinger (KMS) condition \cite{kubo57,martin59}
$g^<(t) = g^<(-t+i\beta \hbar)$. 

\section{\label{secIII}Nonequilibrium Green's Functions, Basic Definitions and Properties}
In this section, we generalize the definitions for the single degree harmonic oscillator and consider a general system described by vibrational displacement $u_j$, where the single index $j$ runs over all the relevant degrees of freedom of the problem.  For example, in a three-dimensional system, $j$ may refer to the $l$-th atom in the $x$ direction.   This compact notation makes the formula valid for any dimensions.  We define the greater Green's function $G^{>}$ as a matrix, with the elements
\begin{equation}
\label{eq-Ggreater-def}
G^{>}_{jk}(t,t') = -\frac{i}{\hbar} {\rm Tr}\bigl[\rho(t_0) u_j(t) u_k(t') \bigr],
\end{equation}
where the trace is the quantum-mechanical trace over a complete set of states, $u_j(t)$ is the Heisenberg operator for the displacement given by
\begin{equation}
\label{eq-heisenberg-operator}
u_j(t) = e^{i(t-t_0)H/\hbar} u_j e^{-i(t-t_0)H/\hbar}, 
\end{equation}
where $u_j$ is the associated Schr\"odinger operator, and $H$ is the Hamiltonian
of the system.   If the Hamiltonian is explicitly time-dependent, one needs to replace the exponential factor by the full Schr\"odinger evolution operator
\begin{equation} 
\label{eq-U-time-order}
U(t,t_0) = T \exp\left( - \frac{i}{\hbar} \int_{t_0}^t H_{t'} dt'\right), \quad (t>t_0),
\end{equation}
i.e., $u_j(t) = U(t_0,t) u_j U(t,t_0)$.  Anti-time order needs to be used if $t<t_0$ in the above formula. We refer to Fetter and Walecka \cite{fetter71}, Chap.~3, for an excellent exposition for the three pictures in quantum mechanics and the properties of the evolution operators.  
The density matrix in Eq.~(\ref{eq-Ggreater-def}) is at time $t_0$.  Since $\rho(t_0)$ here is arbitrary, the system in general is not in equilibrium, and the two-argument function depends on the two times, $t$ and $t'$, separately. By `nonequilibrium', we'll simply mean that $\rho(t_0)$ is not given by a canonical distribution, $\propto e^{-\beta H}$, or the Hamiltonian defining the dynamics may be explicitly time-dependent. 
Note that a reference time, $t_0$, when the Heisenberg picture and Schr\"odinger
picture synchronizes, is arbitrary.  Common choices are either setting
$t_0$ to 0, or the limit $t_0 \to -\infty$.  

Other Green's functions are defined in a similar fashion.  The lesser Green's function can be obtained by swapping time arguments and space indices simultaneously,
\begin{equation}
G^{<}_{jk}(t,t') = G^{>}_{kj}(t',t),
\end{equation}
and the retarded Green's function is obtained by the commutator,
\begin{eqnarray}
G^{r}_{jk}(t,t') &=&  -\frac{i}{\hbar} \theta(t-t') {\rm Tr}\Bigl\{\rho(t_0) \bigl[u_j(t), u_k(t')\bigr] \Bigr\} \nonumber \\
&=& \theta(t-t') \left[ G^{>}_{jk}(t,t') - G^{<}_{jk}(t,t')  \right].
\end{eqnarray}
Similarly, advanced Green's function can be obtained by swapping arguments:
\begin{equation}
G^{a}_{jk}(t,t') =G^{r}_{kj}(t',t).
\end{equation}
The time-ordered and anti-time-ordered Green's functions can be obtained from the others defined above,  $G^t = G^< + G^r$ and $G^{\bar{t}} = G^< - G^a$, as matrix equations. 
We collect some of the useful relations among the Green's functions,
\begin{eqnarray}
\label{eq-Gidentity}
G^t + G^{\bar{t}} &=& G^> + G^<,\\
G^r - G^a &=& G^> - G^<,\\
G^r + G^a &=& G^t - G^{\bar{t}}.
\end{eqnarray}
These linear relations are valid in both the time domain and frequency domain.  In addition, an important relation in the  frequency domain is
\begin{equation}
G^a[\omega] = G^r[\omega]^\dagger,
\end{equation}
where the dagger $\dagger$ stands for hermitian conjugate.  The retarded Green's function is analytic in the upper half plane of the complex frequency domain.  This property guarantees the Kramers-Kronig relation relating the real part of $G^r[\omega]$ with the imaginary part of $G^r[\omega]$ (or vice versa) through Cauchy principle value integrals (we refer to Kubo et al. \cite{kubo-toda-hashitsume92}, Chap.~3.6, or Atland and Simons \cite{altland10}, Chap.~7). 

We have not yet finished with our definitions of Green's functions.  The last and perhaps the most important Green's function in NEGF is the contour-ordered Green's function.  The contour-ordered Green's functions are explained in some books on many-body physics, e.g.,  Haug and Jauho \cite{haug96}, Chap.~4, Zagoskin \cite{zagoskin98}, Chap.~3.4, Rammer \cite{rammer07}, Kleinert \cite{kleinert09}, Chap.~18, or Di Ventra \cite{diventra08}, Chap.~4, and
Kamenev \cite{kamenev11}.
The usefulness of this type of Green's functions is because quantum evolution (the Heisenberg operators and density matrices) is two-sided, see Eq.~(\ref{eq-heisenberg-operator}), where we can think of $u_j(t)$ as developing from the reference time $t_0$ to the time of interest, $t$, using $U(t,t_0)$, meeting the Schr\"odinger operator $u_j$ at time $t$, and then being evolved backward in time by $U(t_0, t)$ from $t$ to $t_0$.  The evolution goes forward and backward, forming a loop, or contour. Another deep reason is that only in this form of contour order, we can develop a transparent perturbation theory, using the interaction picture.

\begin{figure}
\includegraphics[width=\columnwidth]{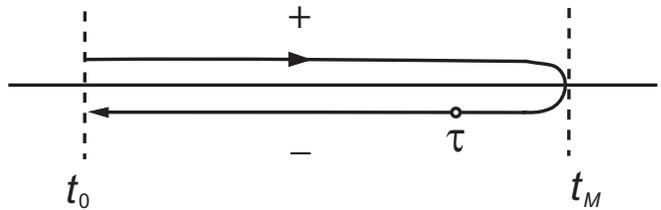}%
\caption{\label{fig:contour}Contour $C$ used to define the nonequilibrium Green's functions.  The upper branch is called $+$ and lower one $-$.  The order follows the direction of the arrows.}
\end{figure}

By convention, we define the contour $C$ as going from $t_0$ in the upper branch (forward going, $+$) or slightly above the real axis on a complex time plane, up to a maximum time $t_M$ relevant to the problem, then returning to the original
time $t_0$ from the lower branch (backward evolving, $-$), or slightly below the real axis, see Fig.~\ref{fig:contour}.  However, the time $t$ is always real, and this has nothing to do with analytic continuation.  We'll use the Greek letter $\tau$ to denote a particular point on the contour and it is equivalent to a time $t$ and a branch index $\sigma = \pm$.  A nice, compact notation for this is $t^\sigma$.  An evolution operator $U$ can be defined on the contour.  If both $\tau_1$ and $\tau_2$ are on the upper branch with $\tau_1$ precedes $\tau_2$ (i.e. $t^+_1 < t^+_2$),  we define
$U(\tau_2, \tau_1) = U(t_2, t_1)$.  This is the usual evolution from a time
$t_1$ to a later time $t_2$ with time ordering [Eq.~(\ref{eq-U-time-order})].
If both $\tau_1$ and $\tau_2$ are on the lower branch with $\tau_1$ precedes $\tau_2$, then $t_1^- > t_2^-$, and $U(\tau_2, \tau_1)$ is given by the same formula
$U(t_2,t_1)$, but since $t_2 < t_1$ the time order should be replaced by anti-time order.  If $\tau_1$ is on the upper branch and $\tau_2$ on the lower branch,
we define $U(\tau_2,\tau_1) = U(t_2,t_M) U(t_M, t_1)$ where the first factor is anti-time ordered, and the second factor is time ordered.  Together, symbolically, we can write
\begin{equation} 
U(\tau_2,\tau_1) = T_c \exp\left( - \frac{i}{\hbar} \int_{\tau_1}^{\tau_2}\!\!\! H_{\tau} d\tau\right), \quad (\tau_2 \succ \tau_1),
\end{equation}
where we assume $\tau_2$ succeeds $\tau_1$ on the contour. $T_c$ is an order super-operator that orders the operators according to a linear order on the contour from earlier to later when read from right to left.  Of course, this makes sense only when the exponential function is expanded as a sum of polynomials of $H_{\tau}$.  The integral is defined as contour integral, which we'll describe further in the next section.  $U$ has a group property on the contour, i.e., if $\tau_3 \succ \tau_2 \succ \tau_1$, then
$U(\tau_3, \tau_2) U(\tau_2, \tau_1) = U(\tau_3, \tau_1)$.  In addition, if $\tau_1 \prec \tau_2$, we define $U(\tau_1, \tau_2)=U(\tau_2,\tau_1)^{-1}$.  With the evolution operator defined on the contour, we can define Heisenberg operator on the contour as
\begin{equation}
\label{eq-O-op}
O(\tau) = U(\tau, t_0^+)^{-1} O\, U(\tau, t_0^+).
\end{equation}
This definition agrees with the usual Heisenberg operator and is independent of the branch index $\sigma$ as an operator acting on a vector of Hilbert space, if the Hamiltonian is independent of branches, which normally is.  However, if $O(\tau)$ is under the contour order sign $T_c$, its position is dictated by the contour variable $\tau$. 

We are now in a position to define the contour ordered Green's function, as
\begin{equation}
\label{G-original-def}
G(\tau, \tau') = - \frac{i}{\hbar} {\rm Tr} \Bigl[ \rho(t_0) T_c u(\tau) u(\tau')^T \Bigr],
\end{equation} 
where $u(\tau)$ now stands for a column vector with $u_j(\tau)$ as elements, and the superscript $T$ stands for matrix transpose, so that $G$ is a square matrix.

Working in the branch component form, $G(\tau, \tau') \to G(t^\sigma,t'^{\sigma'}) = G^{\sigma\sigma'}(t,t')$, we obtain four different Green's functions.  We can identify these Green's functions with the ones defined earlier by comparing the meaning of contour order operator and time-order, anti-time order operator.  If both $\tau$s are on the upper branch, contour order is the same as time order, so we have
$G^{++}=G^{t}$.  Similarly, if both are on the lower branch, contour order is
equivalent to anti-time order, $G^{--}=G^{\bar t}$.  However, if $\tau$ is on the lower branch, and $\tau'$ is on the upper branch, then we don't need to swap positions for all values of $t$ and $t'$ for the operator $u$, so the definition of contour order is equivalent to the greater Green's function, $G^{-+}= G^>$.  Similarly, $G^{+-} = G^<$.  We can write this as a $2\times 2$  matrix  in the space of branches as 
\begin{equation}
G = \left( \begin{array}{cc}
              G^{++} & G^{+-} \\
              G^{-+} & G^{--} 
              \end{array}
     \right) = 
 \left( \begin{array}{cc}
              G^{t} & G^{<} \\
              G^{>} & G^{\bar{t}} 
              \end{array}
     \right).
\end{equation} 

Consider coupled harmonic oscillators (defined by the Hamiltonian $H$, Eq.~(\ref{eq:coupled-oscillator}) below), initially in some mixed state $\rho(t_0)$ for which we assume Wick's theorem is valid. If the system is then driven by an external force, $F(t)$, which is branch dependent, i.e., with an additional $\tau$-dependent external potential $V(\tau) = - F(\tau)^T u(\tau)$, how the density matrix will change?  Schwinger gave a result \cite{schwinger61,argarwalla-pre12} (in our notation)
\begin{eqnarray}
{\rm Tr}\bigl( U(t_M, t_0) \rho(t_0) U(t_0, t_M) \bigr) = \qquad\qquad\qquad\qquad  \\ 
\exp\left( - \frac{i}{2 \hbar} \int_C \int_C F(\tau)^T G(\tau, \tau') F(\tau') d\tau d\tau'\right), \nonumber
\end{eqnarray}  
which motivated him to introduce the $G^{\sigma\sigma'}$.  

\section{Calculus on Contours; convolution, trace, and determinant}
This section is for the mathematically inclined readers.  Those interested in applying
the Green's functions to physics problems can skip this part in the first reading. 
A hallmark of NEGF is the contour valued function. To be able to work with the contour functions, we like to make a few remarks as how differentiation and integration are done on the contour.  Analogous to calculus on the complex plane the derivative on the contour is defined in the usual way,
\begin{equation}
{ d f(\tau) \over d\tau} =
 \lim_{\Delta \tau \to 0} { f(\tau + \Delta \tau) - f(\tau) \over
\Delta \tau},
\end{equation}
where the function $f(\tau)$ is equivalent to two functions, $f^{+}(t)$ and $f^{-}(t)$, for the upper and lower branch, respectively.  On the upper branch, the definition coincides with the usual meaning of derivative with respect to $t$ with $\Delta \tau = \Delta t$.  On the lower branch, the situation is the same, as whether $\Delta \tau$ is positive or negative, it always results as the derivative with respect to $t$.  So symbolically, we say
\begin{equation}
{ d \over d\tau} \to {d \over dt}, \quad\quad { d f(\tau) \over d\tau} \to 
{ d f^{\sigma}(t) \over dt}, \>\sigma = \pm.
\end{equation}
The integration is defined very much like contour integral,
\begin{eqnarray}
\int_{C} d\tau \; &=& \int_{t_0^{+}}^{t_M} dt^{+}\; + 
 \int_{t_M}^{t_0^{-}} dt^{-} \nonumber \\
&=& \sum_{\sigma = \pm 1} \int_{t_0}^{t_M} \sigma dt \;\;.
\end{eqnarray}
The plus and minus signs on $t$ are to make sure the integrand function takes the proper branch indices.  If the integrand is independent of the branches, then the value is zero.

We define the $\theta$ function on contour as
$\theta(\tau, \tau') =1$ if $\tau \succ \tau'$ where $\tau$ succeeds $\tau'$ on the contour, and 0 otherwise.   We define the $\delta$ function by
$\delta(\tau, \tau') = \partial \theta(\tau, \tau') /\partial \tau$.  One can convince oneself that
\begin{equation}
\delta(\tau, \tau') \to \sigma \delta_{\sigma, \sigma'} \delta(t-t'), \quad \sigma, \sigma' = \pm 1,
\end{equation}
where $\delta_{\sigma,\sigma'}$ is the Kronecker delta and $\delta(t-t')$ is the Dirac delta. The $\delta$ function on contour has the expected property that
\begin{equation}
\int \delta(\tau, \tau') f(\tau) d\tau = f(\tau'),
\end{equation}
if a contour contains the point $\tau'$, and 0 otherwise.

In an NEGF calculation, e.g., in collecting terms to form a Dyson equation [e.g., Eq.~(\ref{eq:dyson-G0}) below], one often encounters convolution of a certain type on the contour.  In the theory of full counting statistics, one also needs to evaluate trace or determinant defined on contour.  Part of this is presented in Ref.~\onlinecite{argarwalla-pre12} in appendix C. In the rest of this section, we address these issues, but  first some notations:

`$A$' will mean matrix function with contour times, i.e., $A \to A_{jj'}(\tau, \tau')$. $A(\tau, \tau')$ denotes a matrix with elements $A_{jj'}$ with contour time variables explicitly specified. $A^{\sigma\sigma'}(t,t')$ are the components of $A$.
$\bar{A}^{\sigma\sigma'} \equiv \sigma A^{\sigma\sigma'}$, which has the effect of flipping signs for the bottom two entries of matrix $A$, i.e.,
\begin{equation}
\bar{A} = \left( \begin{array}{cc}
              A^{t} & A^{<} \\
              -A^{>} & -A^{\bar{t}} 
              \end{array}
     \right).
\end{equation} 
$\breve{A}$ is defined as a $45^\circ$ rotation in the space of branches from $\bar A$. With the help of Pauli $z$ matrix, 
\begin{eqnarray}
\sigma_z &=& \left( \begin{array}{cc}
                             1 & 0 \\
                             0 & -1 
                         \end{array} \right), 
\end{eqnarray}
and the rotation matrix
\begin{eqnarray}
R &=& \frac{1}{\sqrt{2}} \left(
\begin{array}{cc}
                             1 & 1 \\
                            -1 & 1 
\end{array} \right), \quad RR^T = I,
\end{eqnarray}
we define
\begin{equation}
\label{eq:k-rotation}
\breve{A} = R^T \sigma_z A R = R^{T} \bar{A} R.
\end{equation}
This is known as a Keldysh rotation (other conventions are also used, e.g., Rammer \cite{rammer07}, Chap.~5.3).  For any $A^{\sigma\sigma'}$, the effect
of the Keldysh rotation is to change to
\begin{eqnarray}
\breve{A} &=& 
\left( \begin{array}{cc}
                             A^r & A^K \\
                             A^{\bar K} & A^a 
\end{array} \right)  \\
 &=&
\frac{1}{2} \left( \begin{array}{cc}
              A^t - A^< + A^> - A^{\bar t}, & A^t + A^{\bar t} + A^< + A^> \\
              A^t + A^{\bar t} - A^< - A^>, & A^< - A^{\bar t} + A^{t} - A^{>}  
                         \end{array} \right). \nonumber
\end{eqnarray}
We should view the above as defining the quantities 
$A^r$, $A^a$, $A^K$, and $A^{\bar{K}}$.  In particular,
$A^{K} \neq A^< + A^>$, as one usually might expect, but
is equal to $(A^t + A^{\bar t} + A^< + A^>)/2$.  We call $A^{K}$ the Keldysh component. Although $A^{\bar{K}}$ need not be
0, it is still true that $A^r - A^a = A^> - A^<$.
For the Green's functions satisfying the relation (\ref{eq-Gidentity})  we get
\begin{equation}
\breve{G} = 
\left( \begin{array}{cc}
                             G^r & G^K \\
                             0 & G^a 
\end{array} \right) .
\end{equation}
The $G^{\bar K}$ component is 0 due to the relation
among Green's functions.  

Convolution of $n$ contour matrix objects is defined as
\begin{eqnarray}
& AB \cdots D& \;\equiv  \nonumber \\
&&\!\!\!\!\!\!\!\!\!\!\!\!\!\!\!\!\!\!\!\!\!\!\! \int d\tau_2 d\tau_3 \cdots d\tau_n 
A(\tau_1, \tau_2) B(\tau_2, \tau_3) \cdots D(\tau_n, \tau_{n+1}),\>\>
\end{eqnarray}
where the usual matrix multiplication in the indices $j$ is implied, and the first and last variables are left free.  So the result is a matrix function of $\tau_1$ and $\tau_{n+1}$. 

We note that matrix equations are invariant under the Keldysh rotation defined by Eq.~(\ref{eq:k-rotation}). In the normal situation when the Green's functions in the Keldysh-rotated space are block-upper triangular, the convolution in real time or product in frequency domain is still upper triangular,
\begin{equation}
\left( \begin{array}{cc}
                             C^r & C^K \\
                             0 & C^a 
\end{array} \right) = 
\left( \begin{array}{cc}
                             A^r & A^K \\
                             0 & A^a 
\end{array} \right) 
\left( \begin{array}{cc}
                             B^r & B^K \\
                             0 & B^a 
\end{array} \right). 
\end{equation} 
Multiplying through the matrices, we find $C^{r} = A^{r}B^{r}$ and similarly for the advanced component, as well as $C^K = A^r B^K + A^K B^a$.  One can also show that $C^{<,>} = A^r B^{<,>} + A^{<,>} B^a$ using the general relations among the Green's functions.  These results are known as Langreth theorem \cite{langreth76}.  Using this technique, it is also fairly easy to find the component form of the Dyson equation, $\breve{G} = \breve{g}  + \breve{g} \breve{\Sigma} \breve{G}$, as 
\begin{eqnarray}
G^r &=& g^r + g^r \Sigma^r G^r, \\
G^K &=& g^K + g^r \Sigma^r G^K + g^r \Sigma^K G^a + g^K \Sigma^a G^a.
\end{eqnarray}
Explicit solutions can be written down as (Ref.~\onlinecite{haug96}, Chaps.~4 and 5)
\begin{eqnarray}
G^r &=& ({(g^r)}^{-1} - \Sigma^r)^{-1},\\
\label{eq:keldysh}
G^< &=& ( 1 + G^r \Sigma^r) g^< (1 + \Sigma^a G^a) + G^r \Sigma^< G^a.
\end{eqnarray}
The last equation above is known as the Keldysh equation. The first term in Eq.~(\ref{eq:keldysh}) is 0 if $(g^{r})^{-1}g^< = 0$.  This is the case for ballistic systems in steady states.

Back to the contour ordered version of functions.
The identity $1$ (in the combined contour time and matrix index space) is defined by the requirement $1A = A$, or $A1B = AB$.  As we can see we can define the identity as $I \delta(\tau, \tau')$ where $I$ is the identity matrix while $\delta$ takes care of the contour space.  The inverse of $A$ is defined by $AB = BA = 1$, $B=A^{-1}$. The inverse has the usual meaning if we represent it in the discretized version of contour time, then $\tilde{A} \tilde{B} = 
\tilde{I}$, $\tilde {B} = \tilde{A}^{-1}$, where the tilded versions are the usual matrices, defined by discretizing the time with a uniform spacing $\Delta t$ and indexed by a triplet of $\sigma$,  $j$, and $t_i (=i \Delta t)$, and  $\tilde{A}_{\sigma jt_i; \sigma' j't_{i'}} = \sigma A_{jj'}^{\sigma\sigma'}(t_i,t_{i'}) \Delta t$.  The tilded version $\tilde{A}$ is useful for numerical computation.
 
A trace on the contour is defined by integrating over all the contour time $\tau_i$ and the usual matrix trace, which can be represented in a number of equivalent ways:
\begin{eqnarray}
{\rm Tr} (A)  &\equiv & \sum_j \int_C  d\tau A_{jj}(\tau, \tau)  \nonumber \\
& = & \sum_{j,\sigma} \int_{t_0}^{t_M} \sigma A_{jj}^{\sigma\sigma}(t,t) dt \nonumber \\
&= & {\rm Tr} ({\bar A}) =  {\rm Tr} (\breve{A}) =  {\rm Tr} ( \tilde{A}).
\end{eqnarray}
Finally, the determinant of $A$ is defined through trace,
\begin{equation}
\det(A) \equiv e^{{\rm Tr}\ln A} = \lim_{\Delta t \to 0} \det (\tilde A).
\end{equation}
With this definition for the determinant, we see that $\det\bigl(I \delta(\tau, \tau')\bigr) = 1$.  If $A$ has the form $1 + M$, we can compute the logarithm of the determinant through a sum of traces, 
\begin{eqnarray}
\ln \det(1+M) &=& {\rm Tr} \ln (1+M) \nonumber \\
&=& {\rm Tr}\left( M - \frac{1}{2} M^2 + \frac{1}{3} M^3 - \cdots \right).\qquad
\end{eqnarray}
This formula has been used to obtain practical numerical method for the computation of cumulants in full counting statistics \cite{argarwalla-pre12}.

\section{Equation of Motion on Contour}
The equation of motion method is a simple and convenient way to get started in an NEGF calculation.  In order to deal with the set of Green's functions defined in real time, which completely characterizes the system, it is useful to consider the equation of motion on the contour $C$ \cite{wang08review}. For a ballistic system, such equation closes, so a complete, exact solution is possible.  However, there are some subtleties (on the boundary conditions/initial conditions) as to how these
equations can be solved \cite{niu99}.  The approach taken here is to express the unknown 
Green's function with what we know, e.g., decoupled equilibrium Green's functions discussed in earlier sections. For interacting systems, the equations become an infinite hierarchy, similar to the Bogoliubov-Born-Green-Kirkwood-Yvon (BBGKY) type of equations.  These hierarchical equations can be put into an integral form which is then equivalent to the Feynman-diagrammatic expansion of the problem.  

The starting point to obtain the equation is the calculus rules outlined in Sec.~IV, and a generalization of the Heisenberg equation of motion on the contour,
\begin{equation}
i \hbar {d O(\tau) \over d \tau } = [O(\tau), H],
\end{equation}
where an arbitrary operator is defined on contour according to Eq.~(\ref{eq-O-op}).  Both the derivative and the operator with the contour variable $\tau$ are equivalent to the ordinary derivative with respect to time and Heisenberg operator at $t$ as far as the effect of operator acting on Hilbert space is concerned.  So the above equation is totally equivalent to the ordinary Heisenberg equation of motion.  The only difference is that when under the contour order sign, $T_c$, the position of the operator needs to be at a proper place ordered according to the contour time $\tau$.  

We illustrate the idea of the equation of motion method with a simple example.  Consider a system of coupled harmonic oscillators, not necessarily in equilibrium, with the
usual Hamiltonian
\begin{equation}
\label{eq:coupled-oscillator}
H = \frac{1}{2} p^T p + \frac{1}{2} u^T K u,\quad K^T = K,
\end{equation}
where $K$ is a symmetric, positive definite spring constant matrix, $u$ is a column vector with component $u_j$, and $p$ is the conjugate momentum vector.  Since the transformation from Schr\"odinger to Heisenberg operator defined on the contour is a unitary transform, the commutation relation holds for the
contour variables at equal time,
\begin{equation}
\label{eq-pq-com}
\bigl[u(\tau), p(\tau)^T\bigr] = i \hbar I,
\end{equation}
where $I$ is an identity matrix having the size equal to the number of degrees of freedom of the problem. 

We write the contour ordered Green's function of the full system in terms of the  $\theta$ function to facilitate easy differentiation,
\begin{eqnarray}
G(\tau, \tau') = -\frac{i}{\hbar} {\rm Tr}\left[ \rho(t_0) T_c u(\tau) u(\tau')^T \right] \qquad\qquad \\
= -\frac{i}{\hbar} \theta(\tau, \tau') \left\langle u(\tau) u(\tau')^T \right\rangle
-\frac{i}{\hbar} \theta(\tau', \tau) \left\langle u(\tau') u(\tau)^T \right\rangle^T. \nonumber
\end{eqnarray}
For notational simplicity, we use angular brackets to denote average over the density matrix, i.e., $ \langle \cdots \rangle = 
{\rm Tr}[ \rho(t_0) \cdots]$.  We now differentiate with respect to $\tau$.  There are two places that depend on $\tau$, one in the $\theta$ function and another inside the average on $u$.  Using
$\partial\theta(\tau, \tau')/\partial\tau = \delta(\tau, \tau')$, 
$\partial\theta(\tau', \tau)/\partial\tau = -\delta(\tau, \tau')$,  the Heisenberg equation $du(\tau)/d\tau = \dot{u}(\tau) = p(\tau)$, and the Leibniz rule, we find 
\begin{eqnarray}
{ \partial G(\tau, \tau') \over \partial \tau}  
&=& -\frac{i}{\hbar} \left\langle T_c \dot{u}(\tau) u(\tau')^T \right\rangle \nonumber \\
&&-\frac{i}{\hbar} \delta(\tau, \tau') \left\langle[ u(\tau), u(\tau')^T] \right\rangle. 
\end{eqnarray}
We have combined the two terms proportional to the $\theta$ functions as a contour ordered one, and combined the two terms of derivatives of the $\theta$ functions as a commutator.  Since $\delta$ is 0 unless $\tau = \tau'$, we can set the second argument to $\tau$.  But equal time coordinates commute, so the second term is 0 [Actually, it is $\infty \cdot 0$, but it is safe to set it to 0].  For phonons, first order equation does not close, hence we take one more derivative to obtain,
\begin{eqnarray}
{ \partial^2 G(\tau, \tau') \over \partial \tau^2}  
&=& -\frac{i}{\hbar} \left\langle T_c \ddot{u}(\tau) u(\tau')^T \right\rangle \nonumber \\
&&-\frac{i}{\hbar} \delta(\tau, \tau') \left\langle[ \dot{u}(\tau), u(\tau')^T] \right\rangle. 
\end{eqnarray}
Commuting the Hamiltonian $H$ with $u$ twice, we obtain the Heisenberg equation $\ddot{u} = -K u$, which has the same form as the classical equation of motion. Further, using the canonical commutation relation, Eq.~(\ref{eq-pq-com}), the second term can be simplified; we obtain the equation of motion of a coupled harmonic oscillator system as
\begin{equation}
\label{eq-G}
{ \partial^2 G(\tau, \tau') \over \partial \tau^2}  + K G(\tau, \tau') = - \delta(\tau, \tau') I. 
\end{equation}

We consider the application of this equation to the problem of thermal transport in a ballistic system.  We create a nonequilibrium but well-controlled situation by partitioning the whole system into three regions, called left lead, center region, and right lead. Each one of the regions will have a well-defined initial density matrix.  Thus the matrix $K$ takes the form
\begin{equation}
K = \left( \begin{array}{ccc}
            K^L & V^{LC} & 0 \\
            V^{CL} & K^C & V^{CR} \\
            0 & V^{RC} & K^{R}
             \end{array}
        \right),
\end{equation}
where the submatrices $K^\alpha$, $\alpha=L,C,R$ are symmetric, and $V^{CL}=(V^{LC})^T$, $V^{CR}=(V^{RC})^T$.  The sizes of the matrices are considered finite at the moment.  If we like to obtain the steady-state result, we'll send the sizes of the leads to infinite at the end of the calculation.
In terms of the Hamiltonians of the subsystems, we may write, 
\begin{equation}
\label{eq-H3part}
H= H_L + H_C +H_R + u_L^T V^{LC} u_C +  u_R^T V^{RC} u_C,
\end{equation}
where the last two terms correspond to the interaction of the leads with the center, and 
the decoupled systems have Hamiltonians, $H_\alpha = (1/2)p_\alpha^T p_\alpha
+ (1/2) u_\alpha^T K^\alpha u_\alpha$, $\alpha = L, C, R$.

We split the $K$ matrix into diagonal and off-diagonal terms as $K = D + {\cal V}$, where
\begin{equation}
D = \left( \begin{array}{ccc}
            K^L & 0 & 0 \\
            0 & K^C & 0 \\
            0 & 0 & K^R
             \end{array}
        \right)\!\!,\>
{\cal V} = \left( \begin{array}{ccc}
            0 & V^{LC} & 0 \\
            V^{CL} & 0 & V^{CR} \\
            0 & V^{RC} & 0
             \end{array}
        \right)\!\!.
\end{equation}
With this split of the decoupled ones and interaction, it is easy to verify that the following Dyson equation is valid,
\begin{equation}
\label{eq-dyson}
G(\tau, \tau') = g(\tau, \tau') + \int_C d\tau'' g(\tau, \tau'') {\cal V} G(\tau'', \tau'),
\end{equation}
where the contour $C$ is the standard contour from $t_0^+$ to $t_M$ and back to $t_0^-$.  We should view the functions $G$ and $g$ strictly defined only in the time interval $[t_0, t_M]$. The advantage of working on this interval instead of the Keldysh open domain $(-\infty, +\infty)$ is that we can treat the transient as well as steady state on an equal footing. The small $g$ given above is defined by
\begin{equation}
\label{eq-small-g}
{ \partial^2 g(\tau, \tau') \over \partial \tau^2}  + D g(\tau, \tau') = - \delta(\tau, \tau') I. 
\end{equation}
Symbolically, we write $G = g + g {\cal V} G = g+G{\cal V}g$, where the multiplication should be understood as a convolution on the contour.
Equation~(\ref{eq-G}) is obtained if we act the differential operator,
$I\partial^2 /\partial \tau^2 + D$, on both sides of Eq.~(\ref{eq-dyson}), using Eq.~(\ref{eq-small-g}) and the property of the $\delta$ function.  The Dyson equation fulfills our goal of expressing the unknown, possibly nonequilibrium Green's function $G$ in terms of simpler Green's function $g$.  

However, it is not a good idea to focus on solving the differential equation (\ref{eq-small-g}) as the solution is not unique.   If $g'$ satisfies $\ddot{g}' + Dg' = 0$, then $g+g'$ also satisfies Eq.~(\ref{eq-small-g}).  Thus, we'll have to fix the small $g$ according to their original definition, Eq.~(\ref{G-original-def}), using the initial density matrices, with the decoupled Hamiltonian $h = (1/2)p^Tp + (1/2) u^T D u = H_L + H_C + H_R$.   Since the Hamiltonian is quadratic, with a product of equilibrium initial states,  the three systems are completely decoupled, we have for $g$ at $\alpha\alpha'$ subblock,
\begin{eqnarray}
g_{\alpha\alpha'}(\tau, \tau') &=& -\frac{i}{\hbar} {\rm Tr}\bigl[ \rho_L(t_0) \rho_C(t_0) \rho_R(t_0) T_c u_\alpha(\tau) u_{\alpha'}(\tau')^T \bigr] \nonumber \\
& = &\delta_{\alpha, \alpha'} g_\alpha(\tau, \tau'), \quad \alpha,\alpha'=L,C,R,
\end{eqnarray}
where we have $(u_L, u_C, u_R)^T=u$, and the time evolution is according to $h$. Obviously, Eq.~(\ref{eq-small-g}) is satisfied with this definition.  If $\rho_\alpha(t_0)$s are the equilibrium distributions, then we have established our goal of `building' the nonequilibrium Green's functions from initially known equilibrium ones.  This will be the usual case, but if the initial state $\rho(t_0)$ is arbitrary, an extra surface term,
$g(\tau, \tau'') \partial G(\tau'', \tau')/\partial \tau''|^{\tau''=t_0^-}_{\tau''=t_0^+} - 
 \partial g(\tau, \tau'')/ \partial\tau'' G(\tau'', \tau')|^{\tau''=t_0^-}_{\tau''=t_0^+}$, needs to be added to the right-hand side of Eq.~(\ref{eq-dyson}).

\section{Feynman Diagrammatics}
The Feynman-diagrammatic perturbation theories are the standard techniques to treat interactions in a systematic way.  We'll refer to the literature for details \cite{abrikosov63,fetter71,doniach74,zagoskin98,mahan00,bruus04,rammer07,altland10}.   Most of the earlier literature treat systems at absolute zero temperature (e.g., quantum field theories).  For finite temperature in thermal equilibrium, it is the Matsubara formalism that is employed.  Fortunately, the diagrammatic structures are all the same, whether it is nonequilibrium contour order, or $T=0$ time order, or Matsubara order.

As an illustration, we study the problem of a ballistic system divided into three regions with the Hamiltonian given by Eq.~(\ref{eq-H3part}) with one additional term for a quartic nonlinear interaction which appears only in the center,
\begin{equation}
\label{eq-Tijkl}
H^n = \frac{1}{4} \sum_{ijkl} T_{ijkl} u_i^C u_j^C u_k^C u_l^C.
\end{equation}
An important step for a perturbative expansion is to separate the system into a solvable one and a perturbation.  A two-step adiabatic switch-on may be used as illustrated in Ref.~\onlinecite{wang08review}.  Here, we'll consider a sudden switch-on at time $t_0$ with the decoupled system $h=H_L + H_C + H_R$ as the unperturbed one and the lead-center
couplings and the nonlinear interactions as perturbation.   In the interaction picture with respect to $h$, the operators and density matrix are transformed unitarily from the Schr\"odinger picture by
\begin{eqnarray}
O_I(t) &=& e^{\frac{i}{\hbar}(t-t_0) h} O  e^{-\frac{i}{\hbar}(t-t_0) h},\\
\rho_I(t) &=&  e^{\frac{i}{\hbar}(t-t_0) h} \rho(t)  e^{-\frac{i}{\hbar}(t-t_0) h},\\
S(t,t') &=&  e^{\frac{i}{\hbar}(t-t_0) h} U(t,t') e^{-\frac{i}{\hbar}(t'-t_0) h} \nonumber \\
&=& T  e^{-\frac{i}{\hbar} \int_{t'}^t \bigl(V_I(t'')+H_I^n(t'')\bigr)dt''},\quad t>t'.
\end{eqnarray}
So the operators follow a `free' evolution of the noninteracting system, and the density matrix evolves through the evolution operator $S$ according to only the interaction part of the Hamiltonian.  Both $O$ and $S$ can be generalized to be defined on the contour (by saying that the time $t$ has an additional contour branch index $\sigma$).

We would like to compute two quantities.  The first is the ``partition'' function or generating function
\begin{eqnarray}
Z &=& {\rm Tr}\bigl[ \rho(t_0) U(t_0, t) U(t,t_0) \bigr] \nonumber \\
&=& 
 {\rm Tr}\left[ \rho(t_0) T_c e^{-\frac{i}{\hbar} \int_C \bigl(V_I(\tau)+H_I^n(\tau)\bigr)d\tau} \right] ,
\end{eqnarray}
where the second line is in the interaction picture with $V_I(\tau) = u_L^I(\tau)^T V^{LC} u_C^I(\tau) + u_R^I(\tau)^T V^{RC} u_C^I(\tau)$ due to the interaction picture transformation.
This quantity $Z$ is clearly 1, by definition.  But we'll take the point of view that the full Hamiltonian may be contour dependent, then $Z = {\rm Tr} [ \rho(t_0) U(t_0^{-}, t_0^{+})]$ may not be 1, a useful point of view when we discuss full counting statistics.  In addition, we consider the functional form of $Z$ in terms of the Green's functions $g$ and look for the relationship between the full Green's functions $G$ and $Z$.  The diagrams generated in $Z$ are known as vacuum diagrams as there are no external lines.   Of course, our main focus is the second quantity, the contour ordered Green's function, 
Eq.~(\ref{G-original-def}).  When transformed into the interaction picture, we have
\begin{eqnarray}
\label{eq-Gaa}
G_{\alpha\alpha'}(\tau, \tau') &=& - \frac{i}{\hbar} {\rm Tr} \Bigl[ \rho(t_0) T_c \bigl\{ u_{\alpha}^I(\tau) u_{\alpha'}^I(\tau')^T  \\
&&\!\!\!\!\!\!\! e^{-\frac{i}{\hbar} \int_C \bigl(V_I(\tau'')+H_I^n(\tau'')\bigr)d\tau''}
\bigr\}\Bigr] {\displaystyle { 1 \over Z}},\quad \alpha = L,C,R.\nonumber
\end{eqnarray}

There are a number of well-known facts or theorems which will be helpful in the development of the Feynman-diagrammatic expansion.  We'll list them here without proofs.  

 1) {\it The Wick theorem}.  This theorem enables one to express the product of terms when the exponential is expanded, in terms of simpler known Green's functions, e.g., 
\begin{eqnarray}
\langle T_c u(1) u(2) u(3) u(4) \rangle \!\!\!\! &= \langle T_c u(1) u(2) \rangle\,
\langle T_c u(3) u(4) \rangle  \nonumber \\
&  + \langle T_c u(1) u(3) \rangle\, \langle T_c u(2) u(4) \rangle\quad \nonumber \\
& + \langle T_c u(1) u(4) \rangle\, \langle T_c u(2) u(3) \rangle, \quad
\end{eqnarray}
where for notational simplicity, we have lumped a set of indices and contour time argument as one single number, e.g., $(1) \equiv (\alpha_1, j_1, \tau_1)$.  
The validity of the Wick theorem relies on the fact that $\ln \rho(t_0)$ is a quadratic form in the dynamic variables. Each graph comes with a numerical prefactor,  which can be found by working out a combinatorial problem of how many ways one can get a topologically equivalent graph due to the Wick decomposition.

\begin{figure}
\includegraphics[width=\columnwidth]{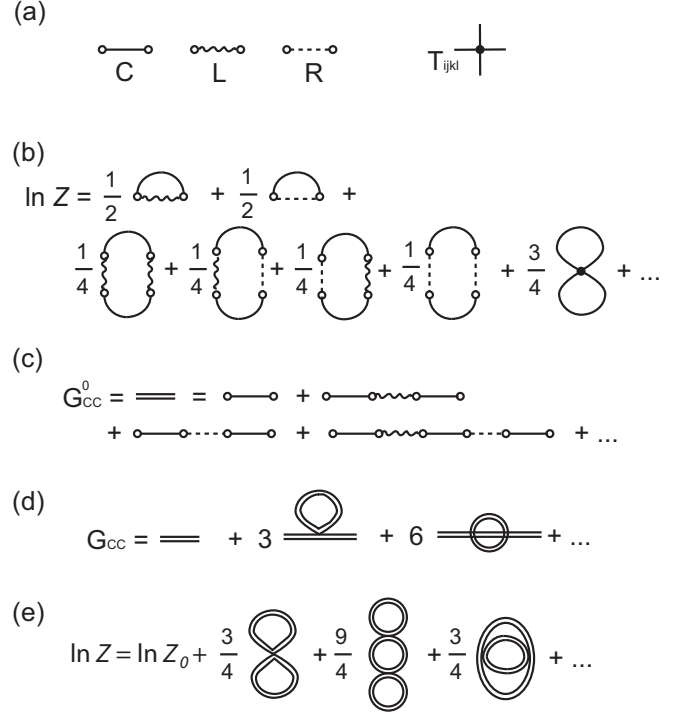}%
\caption{\label{fig:feynman}Feynman diagrams for the nonequilibrium transport problem with quartic nonlinearity. (a) Building blocks of the diagrams.  The solid line is for $g_C$, wavy line for $g_L$, and dash line for $g_R$; (b) first few diagrams for $\ln Z$; (c) Dyson series for the ballistic system Green's function $G^0_{CC}$; (d) Full Green's function $G_{CC}$, and (e) resumed $\ln Z$ where the ballistic result is  $\ln Z_0 = -\frac{1}{2}{\rm Tr}\ln(1 - g_C \Sigma)$.  The number in front of the diagrams represents extra combinatorial factor.}
\end{figure}

To work out the diagrams, for our problem of the center-lead couplings and the quartic nonlinear interaction, we have several building blocks.  First, the pairs of $u$ give the decoupled Green's functions $g_\alpha$, $\alpha = L,C,R$.  This will be drawn as wiggle, straight, and dotted lines, (see Fig.~\ref{fig:feynman}(a)).  These lines are connected possibly in two ways, by the $V^{LC}$ or $V^{RC}$ vertices, which connect the center line with the lead lines, and by $T_{ijkl}$ which connects four center lines.  For $G$ we have two external terminals labeled 1 and 2.  For $Z$ all variables are dummy, and need to be summed.  

2) {\it Cluster decomposition theorem, factor theorem}.  The graphs of $Z$ contain connected and disconnected pieces.  The cluster decomposition theorem is a very general theorem which says  that if we take the logarithm, then $\ln Z$ is given only by the connected graphs (see Fig.~\ref{fig:feynman}(b)).  A similar statement holds in the Mayer's cluster expansion in equilibrium statistical mechanics for interacting gases (Friedman \cite{friedman85}, Chap.~6).   In addition, the disconnected pieces do not enter into the diagrams for the Green's functions
$G$.  This can be understood in two ways, first is that due to the denominator $Z$ in Eq.~(\ref{eq-Gaa}),  the disconnected pieces (which contains only vacuum diagrams) get exactly cancelled.  Alternatively, if Eq.~(\ref{eq-Gidentity}) holds, then all the vacuum diagrams are numerically 0, so such diagrams do not appear and $Z=1$ \cite{bijay-thesis}. 

3) {\it The Dyson equation}.  Certain diagrams can be regrouped and easily summed.  Let us first consider the case where the nonlinear interaction vanishes.  Then the diagrams for the corresponding Green's function $G_{CC}^0$ can only be a linear chain, with  binomial combinatorial ways of putting the left or right lead lines, see Fig.~\ref{fig:feynman}(c).  If we define the self-energies as
\begin{equation}
\label{eq-self-energy-def}
\Sigma_{\alpha}(\tau,\tau') = V^{C\alpha} g_{\alpha}(\tau, \tau') V^{\alpha C}, \quad \alpha =L, R,
\end{equation}
the terms can be expressed in a recursive way.  We thus have (for the center part of the full ballistic Green's function)
\begin{eqnarray}
G_{CC}^0 &=& g_C + g_C \Sigma g_C + g_C \Sigma g_C \Sigma g_C + \cdots \nonumber \\
\label{eq:dyson-G0}
&=&  g_C + g_C \Sigma G_{CC}^0,
\end{eqnarray}
where $\Sigma = \Sigma_L +\Sigma_R$ and convolution on the contour is implied.
This is the Dyson equation for the central region, which has a similar mathematical structure as Eq.~(\ref{eq-dyson}).

The nonlinear part is more complex, but in terms of $G_{CC}^0$ (double line in the graphs), the diagrams for $\ln Z$ and full nonlinear Green's function $G_{CC}$ can be simplified, as shown in Fig.~\ref{fig:feynman}(e,d).  Finally, we can define the nonlinear self-energy as part of the diagrams where it is not singly connected (double or more connectivity) and with the two external legs chopped, thus giving
\begin{equation}
\label{eq:dyson-Gsn}
G_{CC} = G_{CC}^0 + G_{CC}^0 \Sigma_n G_{CC}.
\end{equation}

4) One can introduce a vertex function (and Hedin-like equation) and encapsulate the diagrams more compactly using functional derivatives \cite{kwok68,rammer07,mlleek12}. How useful they are for practical calculation remains to be seen. 

5) {\it Connection between vacuum diagrams and Green's function}.  This fact seems less well-known.  We notice that the vacuum diagrams given in $\ln Z$ and the graphs formed by ${\rm Tr}( G_{CC} \Sigma)$ are the same, where the trace means both for the space index $j$ and contour time $\tau$, and the expression means to close the two external lines with one more self-energy line.  However, the combinatorial prefactors differ.  This difference can be removed if we differentiate with respect to the self-energy $\Sigma$.  Thus we have the following identity,
\begin{equation}
\label{eq-dZG}
\delta \ln Z = \frac{1}{2} {\rm Tr}\left( G_{CC} \delta \Sigma \right),
\end{equation}
where the variation $\delta$ means the functional form of the self-energy is varied while $g_{C}$ holds constant.  This relation can be derived in a more rigorous, algebraic way \cite{kleinert-pre-00}.
 
\section{Landauer Formula}
So far we have studied the properties of the Green's functions and how such functions can be calculated for general linear or nonlinear systems.  In this section, we look at one of the most important physical observables in transport, i.e., the thermal or energy current.  The energy current transported out of the left lead is defined as 
\begin{equation}
I_L(t) = -\left\langle { d H_L(t) \over dt }\right\rangle =  \bigl\langle \dot{u}_L(t)^T V^{LC} u_C(t) \bigr\rangle,
\end{equation} 
where the angular brackets denote trace over the initial density matrix $\rho(t_0)$ and the operators are in Heisenberg picture at time $t$.
This energy (per second) is presumably transferred to the center or the coupling between the left lead and the center, since the energy of the whole system is conserved and the left lead is connected directly to the center but not to the right lead.

We need to connect the definition of the current with the Green's functions.
Using the definition of $G_{CL}^<$ or $G_{CL}^>$ (or $G^K_{CL}/2$) we can write
\begin{equation}
\label{eq-ILdef2}
I_L(t) = i \hbar { \partial \; \over \partial t' } {\rm Tr}\bigl[ G_{CL}^{<,>}(t,t') V^{LC} \bigr]\Big|_{t'=t}.
\end{equation}
The trace above is in the sense of an ordinary matrix trace by summing over the diagonal elements. 
The Green's functions above use the mixed lead and center degrees of freedom.  Observing the fact that centers are usually more complex but finite, and leads are simple (free phonons) but may be infinite, we can try to relate $G_{CC}$ to $G_{CL}$ 
using the Dyson equation, (\ref{eq-dyson}), in the form $G = g + G{\cal V}g$.    Working out the $CL$ component of this block matrix equation, we obtain
$G_{CL} = G_{CC} V^{CL} g_L$, or in full detail
\begin{equation}
G_{CL}(\tau, \tau') = \int_C G_{CC}(\tau, \tau'') V^{CL} g_L(\tau'', \tau') d\tau''.
\end{equation}
Although we obtained this equation from the ballistic Dyson equation, the specific properties of the center has not been used.  It turns out that this equation is also valid if the center is nonlinear.  This can be shown by looking at the equation of the motion of $G_{CL}$ directly.  Using the Langreth theorem for the $G_{CL}^<$ component and substituting the result into Eq.~(\ref{eq-ILdef2}), we obtain
\begin{eqnarray}
I_L(t) &=& i \hbar {\rm Tr} \int_{t_0}^t \Bigl[ G_{CC}^{r}(t,t'') { \partial \; \over \partial t' }
\Sigma_L^{<}(t'',t')\big|_{t'=t} \nonumber \\
&& \qquad +\; G_{CC}^{<}(t,t'') 
{ \partial \; \over \partial t' }\Sigma_L^{a}(t'',t')\big|_{t'=t} \Bigr] dt'',
\end{eqnarray}
where we have used the definition of lead self-energy, Eq.~(\ref{eq-self-energy-def}). 

The above formula is valid for any time $t$.  If steady state result is 
required, we can send $t_0 \to -\infty$, and perform a Fourier transform and after applying the convolution theorem for Fourier transform, we obtain
\begin{eqnarray}
I_L &=& - \int_{-\infty}^{+\infty}\!\!\! { d \omega \over 2 \pi} \hbar \omega 
{\rm Tr} \Bigl[ G^r[\omega] \Sigma_L^<[\omega] +
G^<[\omega] \Sigma_L^a[\omega] \Bigr]  \nonumber \\
&=&  \int_{-\infty}^{+\infty}\!\!\! { d \omega \over 4 \pi} \hbar \omega\, 
{\rm Tr} \Bigl[ G^< \Sigma_L^> -
G^> \Sigma_L^< \Bigr] .
\end{eqnarray} 
For simplicity, we have dropped the $CC$ subscript on $G$; we have also omitted the $\omega$ arguments in the second line.  The last equation can be obtained by taking $(I_L+I_L^*)/2$ since $I_L$ must be real, and using the relations among Green's functions discussed in Sec.~\ref{secIII}.
The above formula is known as the Meir-Wingreen formula first derived for  electronic transport \cite{meir92}. 

The Meir-Wingreen formula is valid for ballistic as well as interacting centers.  If the center is ballistic, i.e., $\Sigma_n=0$ in Eq.~(\ref{eq:dyson-Gsn}), the formula can be further simplified.  The result for the ballistic system is called Landauer formula with a transmission function $T(\omega)={\rm Tr}\bigl(G^r \Gamma_L G^a \Gamma_R\bigr)$ known as the Caroli formula \cite{caroli71}.  Thus, using the definition for the lead spectral function
\begin{equation}
\Gamma_{\alpha} = i \bigl(\Sigma_\alpha^r - \Sigma_\alpha^a),\quad \alpha =L, R,
\end{equation}
and the relations (as a consequence of the Dyson equation for the center, see, e.g., Datta \cite{dattabook1}, Chap.~8)
\begin{eqnarray}
G^<=G^r (\Sigma_L^< + \Sigma^<_R) G^a,\\
G^a - G^r = i G^r (\Gamma_L + \Gamma_R) G^a,
\end{eqnarray}
we finally obtain
\begin{equation}
\label{eq-landauer-like}
I_L = \int_0^{+\infty}\!\! {d\omega \over 2 \pi} \hbar \omega\, T(\omega) (f_L - f_R),
\end{equation}
where $f_\alpha = 1/(e^{\beta_\alpha \hbar \omega} - 1)$, $\alpha = L, R$, is the Bose-Einstein distribution for the leads.   A very detailed
derivation of the above is given in Leek \cite{mlleek12}, Chap.~3.
A similar formula for electrons was first given by Landauer \cite{landauer57,landauer70} from a wave scattering point of view.  It is
appropriate to call Eq.~(\ref{eq-landauer-like}) Landauer-like formula and it has been derived in a number of different ways for thermal transport \cite{ozpineci01,segal03,mingo03,dhar06,dharprb06,jswprb06,yamamoto06,wzhang07,das-dhar12}.

As it is seen that the surface Green's functions $g_L$ and $g_R$ are the important inputs for the
nonequilibrium transport problems, it is required to develop algorithms to calculate them efficiently for realistic systems.  Fortunately, this has already been done quite early \cite{sancho84,sancho85} for electron transport, but it applies equally well for phononic systems. Algorithmic procedure for the calculation of the surface Green's function and thus the self-energies are reviewed in Ref.~\onlinecite{wang08review}. 

In the rest of this section, we give a simple example of calculating the Green's functions and transmission coefficient for a uniform one-dimensional chain with the force constant matrix $K$ which is $2k+k_0$ along the diagonal and $-k$ along the two first off-diagonals.   We split the system into three regions with $N_L$, $N_C$, and $N_R$ number of particles for each.  The first step in such a calculation is to determine the surface Green's functions.  The eigenvalues and eigenvectors of the uniform tridiagonal matrix can be  obtained analytically \cite{arya90,cuansing12}, given
\begin{eqnarray}
\Omega^2_n &=& 2k\bigl(1-\cos(q_n)\bigr) + k_0, \quad q_n = \frac{\pi n}{N+1},  \\
u_j^n &=& \sqrt{ \frac{2}{N+1} } \sin(q_n j),\quad n = 1, 2, \cdots, N,
\end{eqnarray}
where $N$ can be one of the $N_\alpha$, $\alpha = L, C, R$. 
We construct an orthogonal matrix $S$, $S^T S = I$, by $S=(u^1, u^2, \cdots, u^N)$ such that $S^T K S = {\rm diag}\bigl( \Omega_1^2, \Omega_2^2, \cdots, \Omega_N^2 \bigr)$.  Each mode follows the results of the single degree harmonic oscillator discussed in Sec.~II. Then the retarded Green's function in the time domain for a chain of $N$ sites is  
\begin{equation}
\label{eq:finitegr}
g^r(t) = S\, {\rm diag} \left\{ - \theta(t) \frac{ \sin(\Omega_j t)}{\Omega_j} \right\} S^T.
\end{equation}
For steady state, we need to have an infinite lead, $N \to \infty$.  In order to take this limit, it is more convenient to solve the retarded Green's function in the frequency domain,
\begin{equation}
\bigl( (\omega  + i \eta)^2 - K^{\alpha} \bigr) g^r_{\alpha}[\omega] = I,\quad \alpha = L,C,R.
\end{equation}
The (1,1) element can be found analytically, given 
\begin{equation}
g^r[\omega]_{11} = - \frac{\lambda}{k} \, \frac{ 1 - \lambda^{2N}}{1 - \lambda^{2N + 2}},
\end{equation}
where $\lambda$ is the solution of the quadratic equation,
\begin{equation}
k\lambda^{-1} + (\omega + i \eta)^2 -2k -k_0 + k \lambda = 0,
\end{equation}
with $| \lambda | < 1$  (the small imaginary number $i \eta$ makes this choice unambiguous).
Since $| \lambda | < 1$, in the limit $N \to \infty$, we obtain a simple result for the surface Green's function of the semi-infinite lead as $g^r[\omega]_{11} = - \lambda/k$.
Using the definition of self-energy, Eq.~(\ref{eq-self-energy-def}), we obtain for the matrix elements $\Sigma_L^r[\omega]_{11} = \Sigma_R^r[\omega]_{N_{C}N_{C}} = -k \lambda$,  and 0 for all other elements.

The retarded Green's function of the coupled system in the center can be obtained by solving the Dyson equation, Eq.~(\ref{eq:dyson-G0}), of the retarded component, using Langreth theorem and in the frequency domain, 
\begin{equation}
G^r = g_C^r + g_C^r \Sigma^r G^r.
\end{equation}  
However,  we can also obtain $G^r$ by considering a similar equation as for $g$ for the whole space domain, with integer index $j,l$ vary from $-\infty$ to $+\infty$, i.e.,
\begin{equation}
kG^r_{j-1,l} + \bigl((\omega + i \eta)^2 -2k -k_0\bigr)G_{j,l} + k G^r_{j+1,l} = \delta_{j,l}.
\end{equation}
Since $G^r$ must be translationally invariant in space indices, and must decay to 0 when $j$ or $l \to \pm \infty$, we have \cite{jswpre07} $G_{jl}[\omega] = c \lambda^{|j-l|}$, where $c = 1/[(\lambda - 1/\lambda)k]$ is fixed by the $j=l$ diagonal equation. 
Finally, the transmission coefficient is found by the Caroli formula.  After some algebra, one finds $T(\omega) = 1$ if $\sqrt{k_0} < \omega < \sqrt{4k+k_0}$ and 0 otherwise. Of course, this simple result is expected if one thinks of it from a wave scattering picture \cite{wangj06prb,lfzhang-keblinski11} without doing any calculation.

\section{Multiple Leads, Lead-Lead Interaction}
In this section, we give some formulas without much derivation for ballistic transport in some more settings.  The first is that of multiple leads.  This is much in parallel to B\"uttiker's theory \cite{buttiker86,buttiker88a,buttiker88b,dattabook1,blanter00} for electron transport with multiple leads.   We define 
\begin{equation}
T_{\alpha\alpha'}(\omega) = {\rm Tr}\bigl( G^r \Gamma_\alpha G^a \Gamma_{\alpha'} \bigr)
\end{equation}
for the transmission coefficient from the $\alpha$ to $\alpha'$ lead.  Then the current out of the $\alpha$ lead is given by summing over contributions from all the other leads,
\begin{equation}
I_\alpha = \int_0^{+\infty}\!\! {d\omega \over 2 \pi} \hbar \omega\, \sum_{\alpha' \neq \alpha} T_{{\alpha'}\alpha}(\omega) (f_\alpha - f_{\alpha'}).
\end{equation}
Three-terminal problems are studied in Ref.~\onlinecite{lfzhang-prb-10} and \onlinecite{zxxie12} where one of the terminals is treated as a B\"uttiker probe, i.e., the third terminal is required to have zero current, determined self-consistently by adjusting its bath temperature.  This mimics an inelastic scattering, thus resulting in thermal rectification even for ballistic systems, as well as diffusive transport for long chains \cite{dhar-roy-jsp-06,roypre08,bandyopadhyay11} (where each atom gets a probe or self-consistent reservoir).   However, the effective decoherence is a bit artificial, and its relevance to truly nonlinear systems is not clear.
A four-terminal problem is treated in Ref.~\onlinecite{lfzhang-njp-09} for the spin-phonon or phonon Hall interaction (still a ballistic problem since the ``interaction'' is bilinear in coordinates and momenta).

Now we come back to the two-lead problem again.  In the standard modeling of such systems, one always assumes that there is no interaction between the left lead and the right lead.  This, of course, can be achieved if the center region is large enough and the interactions are short-ranged.  However, in practical calculation, one always finds a small residue of the left-right lead interaction.  Now the question is, can we have a generalization of the Caroli formula so that the left-right lead interaction is allowed?
The answer turns out yes, with a new formula \cite{hnlee12}:
\begin{equation}
T_g(\omega) = {\rm Tr} \bigl( G^a_{RL} \tilde{\Gamma}_L  G_{LR}^r \tilde\Gamma_R \bigr),
\end{equation}
where the new tilded lead spectral function is
\begin{equation}
\tilde \Gamma_{\alpha} = i \left[ \left(g^a_{{\rm red},\alpha}\right)^{-1}
-  \left(g^r_{{\rm red},\alpha}\right)^{-1} \right],\quad \alpha =L, R.
\end{equation}
The subscript \emph{red} indicates that the surface Green's functions are not the full one but reduced subblock large enough so that outside that sizes, the left-right couplings (as well as center-left, center-right) are 0.  Also, the matrix $G^r_{LR}$ is not infinite large but a finite piece consistent with the sizes of $g_{{\rm red},\alpha}$.
The retarded Green's function in the $LR$ subblock can be obtained by solving a Dyson equation
\begin{eqnarray}
G^r_{LR} &=& \tilde{g}^r_L \tilde{V}^{LR} \tilde{g}^r_R + 
\tilde{g}^r_L \tilde{V}^{LR} \tilde{g}^r_R \tilde{V}^{RL} G^r_{LR},\\
\tilde{g}^r_\alpha &=& \left[ \left(g^r_{{\rm red},\alpha}\right)^{-1} - V^{\alpha C} g^r_C V^{C \alpha} \right]^{-1}\!\!\!\!,\quad \alpha = L, R, \nonumber \\
\left(\tilde{V}^{RL}\right)^T &=& \tilde{V}^{LR} = V^{LR} + V^{LC} g_C^r V^{CR},
\end{eqnarray}
where $V^{LR}=(V^{RL})^T$ is the left-right lead coupling matrix
and $G^a_{RL} = (G^r_{LR})^\dagger$.   It is also possible to give an expression
as a special case of the above for the transmission where the system has no center and left and right leads are directly connected (an interface problem),
as 
\begin{equation}
T_{{\rm i}}(\omega) = {\rm Tr} \bigl( G^r_{RR} \tilde{\Gamma}_R G^a_{RR} \gamma_L \bigr),
\end{equation}
where $\gamma_L = i V^{RL}(g^r_L - g^a_L)V^{LR}$.
The starting point of deriving these results is Eq.~(\ref{eq-dyson}), where ${\cal V}$ is non-zero for all the subblocks except on the diagonal. We refer to Ref.~\onlinecite{hnlee12} for more details.

\section{Full counting statistics}
In thermal transport, the current is the first and most basic quantity to look at.  However, other related quantities are also relevant and important.  One of them is the current fluctuation.   For example, we may consider the time-displaced current-current correlation function (much work has been done for electron shot noise \cite{blanter00}).   In a transient situation, the fluctuations are large.  Due to the stochastic nature of the baths, each individual experiment will give a different result $Q$ ($=Q_L$) if we measure the amount of energy transferred in a fixed amount of time $t_M$ out of the left lead.  Thus, it is interesting and useful to look at the distribution of the energies.  This distribution satisfies certain `fluctuation theorem' under certain conditions (e.g., long time), which is now a very hot area of research \cite{esposito-RMP09,campisi11,seifert1205.4176}. The complete distribution, or equivalently the moment generating function (characteristic function),
\begin{equation}
Z(\xi) = \langle e^{i\xi Q} \rangle = \int_{-\infty}^{+\infty}\!\!\! dQ\, e^{i\xi Q} P(Q),
\end{equation}  
reveals more about the system, especially its quantum nature.  
When the function $Z(\xi)$ is known, the moments of $Q$ can be computed by taking derivatives, $\langle Q^n \rangle = \partial^n Z/\partial (i\xi)^n$, and then setting $\xi=0$, and the cumulants are defined through  $\langle\langle Q^n \rangle\rangle = \partial^n \ln Z/\partial (i\xi)^n|_{\xi=0}$.  In particular, the first moment or cumulant is proportional to the current in long time, $\langle Q \rangle = \langle \langle Q \rangle\rangle \approx t_MI$; the second cumulant is the variance
$\langle \langle Q^2 \rangle\rangle = \langle Q^2 \rangle - \langle Q \rangle^2$, and so on.
This problem of the study of the distribution of the transferred quantity is known as
full counting statistics in the electronic transport literature.  There, the number of electrons transferred in a given time is a discrete quantity, thus the word `counting' is appropriate.   Phonons cannot be counted (however, see Ref.~\onlinecite{roukes99}), and also we are not interested in the number of  phonons since it is not a conserved quantity.  What we do here is to measure the amount of energy, a continuous variable,  transferred from the left lead into the center.   In statistical mechanics literature, $Z$ is related to so-called large deviation problem  (when time $t_M$ approaches infinity) \cite{touchette09}.  

In defining the generating function,  
the most important observation is that the energy transferred, $Q$, is not associated with the eigenvalues of a quantum-mechanical operator.  Instead, it is computed by the difference of the energies of the left lead at two different times, $Q = a-b$, where
$a$ and $b$ are the eigenvalues of $H_L$  at time $t_0$ and $t_M$, respectively.  Using this two-time measurement protocol and the standard von Neumann's interpretation of quantum measurement, we can derive a very general formula for $Z(\xi)$ using product initial state given by \cite{esposito-RMP09,lihuanan-prb12},
\begin{eqnarray}
Z(\xi) &=& \sum_{a,b} e^{i\xi(a-b)} P(b,a) \\
&=& {\rm Tr}\left[ \rho_L(t_0)\rho_C(t_0)\rho_R(t_0) e^{i\xi H_L} e^{-i\xi H_L(t_M)} \right] \nonumber
\end{eqnarray}  
where $P(b,a)$ is the joint distribution for the event history 
which at the initial measurement resulted in the left lead with energy
$a$ and $b$ at the second measurement at time $t_M$,
assuming discrete energy spectrum.   If the initial state is not a product of equilibrium states, the situation is more complicated.  We refer to Ref.~\onlinecite{argarwalla-pre12} for more details. 

We can write $Z$ in terms of the modified evolution operator $U_{x}(t,t')$ governed by a modified Hamiltonian $H_x =e^{ixH_L} H e^{-ixH_L}$ which is contour branch dependent with $x=-\xi/2$ on the upper (forward) branch and $x=\xi/2$ on the lower (return) branch, giving $Z = \langle U_{\xi/2}(t_0,t_M) U_{-\xi/2}(t_M,t_0) \rangle$.  Transforming into the interaction picture with respect to the decoupled system $h=H_L + H_C + H_R$, we obtain
\begin{equation}
Z(\xi) = \langle T_c e^{- \frac{i}{\hbar} \int_C V^x(\tau) d\tau} \rangle,
\end{equation}
where $V^x = u^{x,T}_L V^{LC}u_C + u_C^T V^{CR}u_R$ in the ballistic case.  The effect of measurement is to replace the variable $u_L$ by a transformed one.  In the interaction picture, this is equivalent to shifting  time argument, i.e.,
$u^{x}(\tau) = u\bigl(\tau + \hbar x(\tau)\bigr)$ where the amount of shift depends on the branch.  It is $-\hbar \xi/2$ on the upper branch and $\hbar \xi/2$ on the lower branch.  
We then use the Feynman-diagrammatic method to expand the exponential and group the various terms.  After some simplification, we obtain \cite{wang-prb11,argarwalla-pre12} 
\begin{equation}
\label{eq-Z1}
\ln Z(\xi) = - \frac{1}{2} {\rm Tr}_{j,\tau} \ln \bigl(1 - G^0_{CC} \Sigma^A_L\bigr),
\end{equation}
where the trace is over the ordinary space index $j$ as well as over the
contour time $\tau$, $G^0_{CC}$ is the standard ballistic contour ordered Green's function of the center, while the important new self-energy 
\begin{equation}
\Sigma_L^A(\tau, \tau') = \Sigma_L\bigl(\tau + \hbar x(\tau), \tau' + \hbar x(\tau')\bigr) - 
\Sigma_L(\tau, \tau') 
\end{equation} 
is the difference between the left-lead argument-shifted self-energy and the original standard lead self-energy.
An alternative expression valid also for interacting systems \cite{gogolin06,lihuanan-1210.2798} for the derivative of $\ln Z$ with respect to $i\xi$ is [using Eq.~(\ref{eq-dZG})],
\begin{equation}
\label{eq-lnZ2}
{ \partial \ln Z \over \partial (i\xi) } = 
\frac{1}{2} \int_C \!d\tau \int_C\! d\tau' {\rm Tr} \left[ \tilde{G}_{CC}(\tau, \tau') 
{ \partial \tilde{\Sigma}_L(\tau',\tau) \over \partial (i\xi)}   \right].
\end{equation}
We can think of this equation as a generalization of the Meir-Wingreen current formula to the full counting statistics.  The meaning of the tildes there is that we express all quantities in terms of small $g_\alpha$, $\alpha=L,C,R$, and then replace all occurrence of $g_L$ by
\begin{equation}
\tilde{g}_L(\tau, \tau')  = - \frac{i}{\hbar} \langle T_c u_L^x(\tau) u_L^x(\tau')^T \rangle,
\end{equation}
which is simply an argument-shifted version of the original left lead Green's function.

We note that Eq.~(\ref{eq-Z1}) or (\ref{eq-lnZ2}) is defined on the segment of the contour $C$, so the result is valid both for transient and steady state (if we take $t_0 \to -\infty$ and $t_M \to + \infty$).  In the long-time limit, after transforming the Green's functions into frequency domain using the property of time-translational invariance, and employing the standard relations of Green's function and some algebra, the cumulant generating function for a ballistic, left-center-right junction system is then given as
\begin{eqnarray}
\label{eq:lnZ-longtime}
\ln Z(\xi)= - t_M \int_{-\infty}^{+\infty}\! \frac{d\omega}{4\pi} \ln \det \Bigg\{
I - G_0^r \Gamma_L G^a_0 \Gamma_R \Big[ \qquad\nonumber \\
 f_L(1+f_R) (e^{i\xi \hbar \omega} - 1)  + f_R( 1 + f_L) (e^{-i\xi \hbar \omega}-1)   \Big] \Bigg\}.\quad\quad
\end{eqnarray}
This is the phonon analog of the famous Levitov-Lesovik formula for electrons \cite{levitov93,levitov96}, first given by Saito and Dhar \cite{saito-prl07}.  The long-time generating function satisfies the relation $Z(\xi) = Z\bigl(-\xi + i(\beta_R - \beta_L)\bigr)$, which is a form of the Gallavotti-Cohen symmetry \cite{gallavotti95}. 
Classical versions are given in Refs.~\onlinecite{kundu11,saito11}.
Similar generating functions have also been obtained for systems with driven forces \cite{argarwalla-pre12}, with the left-right interaction term, $u_L^T V^{LR}u_R$ \cite{lihuanan-prb12}, as well as an extension to nonlinear systems starting from Eq.~(\ref{eq-dZG}) through a counting field $\xi$-dependent nonlinear self-energy $\tilde{\Sigma}_n$ \cite{lihuanan-1210.2798}.  Further details can be found in
Ref.~\onlinecite{bijay-thesis}.

\begin{figure}
\includegraphics[width=\columnwidth]{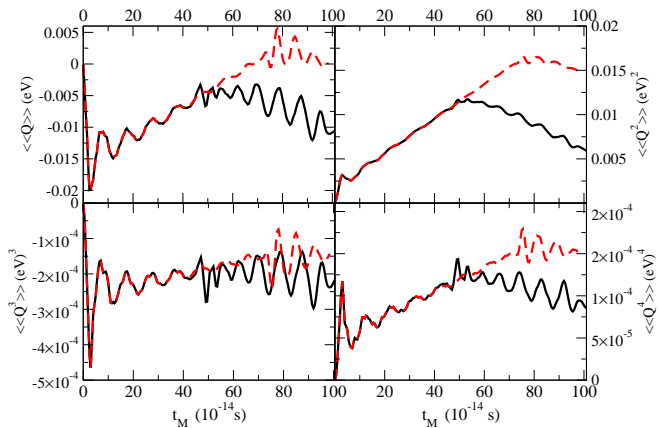}
\caption{\label{fig:bijay}(Color online) Plot of the cumulants of heat $\langle \langle Q^{n} \rangle \rangle$
for $n=1, 2, 3$, and 4 for one-atom center connected with two finite leads (one-dimensional chain) as a function of measurement time $t_M$. 
The black (solid) and red (dashed) line correspond to $N_L = N_R = 20$ and $N_L = N_R = 30$,
respectively. The initial temperatures of the left, center, and the right parts are $310\,$K, $360\,$K, and
$290\,$K, respectively. We choose $k=1$\,eV/(u\AA$^2$) and $k_0=0.1$\, eV/(u\AA$^2$) for all particles.}
\end{figure}

We present a numerical example applying the theory outlined above for a uniform one-dimensional chain. 
In Fig.~\ref{fig:bijay} we plot the first four cumulants of energy transferred for ballistic system with one atom at the center which is connected with two finite-size leads. We use the formula given in Eq.~(\ref{eq-Z1}) and numerically evaluate the derivatives of $\ln Z(\xi)$ with respect to the counting parameter $\xi$. The self-energy $\Sigma_L^A$ for the finite lead is calculated using the lesser version of Eq.~(\ref{eq:finitegr}). $G_{CC}^{0}$ is obtained by numerically solving the Dyson equation given in Eq.~(\ref{eq:dyson-G0}). The plot shows that for finite leads all the cumulants reaches a quasi-steady state with a finite recurrence time $t_r$ which depends on the length of the full system and the velocity of the phonon waves. After $t_r$ phonon waves which are scattered back from the boundaries interfere and this results in the cumulants to oscillate rapidly.  Similar results are obtained for a left-right lead problem without center \cite{cuansing12}.  For infinite size leads \cite{cuansing10} complete irreversible behavior emerges and the system achieves a unique steady state with infinite recurrence time. The slopes in the quasi-steady state regime match with the predicted values obtained from Eq.~(\ref{eq:lnZ-longtime}).

\section{Nonlinear Systems, Perturbation - thermal expansion, phonon life time}

NEGF offers a straightforward treatment for a perturbative expansion result when systems are nonlinear.  We illustrate this with two applications: the problem of thermal expansion and phonon life time.  The thermal expansion in bulk systems is usually treated with the standard Gr\"uneisen theory where the size-dependent vibrational frequencies (through the Gr\"uneisen parameter) are the key parameters, see, Ashcroft and Mermin \cite{ashcroft-mermin76}, Chap.~25.   For a finite system, we look directly at the equilibrium displacements with a proper boundary condition (e.g., one side of a graphene sheet is fixed) in comparison with a corresponding ballistic one due to the lowest order nonlinear effect \cite{jiang-prb80-205429}.   To this end, it is useful to introduce a one-point contour ordered Green's function,
\begin{equation}
G_j(\tau) = - \frac{i}{\hbar} \langle T_c u_j(\tau) \rangle.
\end{equation}
The contour order does not play any role here, but it is convenient and uniform in notation when we relate the one-point Green's function to the usual two-point one.  To lowest order of the $T_{ijk}$ cubic nonlinearity, there is only one diagram (Feynman diagrams up to second order in $\hbar$ are given in Ref.~\onlinecite{jiang-prb80-205429}).  We call it the lollipop diagram and is given, algebraically,
\begin{equation}
G_j(\tau) = \sum_{lmn} T_{lmn} \int \!d\tau'\, G^0_{lm}(\tau',\tau')G^0_{nj}(\tau',\tau),
\end{equation}
where the superscript 0 refers to the unperturbed, ballistic system Green's functions.
In equilibrium or steady state, the contour is from $-\infty$ to $+\infty$ and back to $-\infty$.  Thus, the result is independent of time and branch index.  The Green's function can be further expressed in real time or in frequency domain using the Langreth rules (e.g.,
$G^{<}(0) \int dt\, G^r(t)$).  The displacement relative to the ballistic equilibrium (which gives $\langle u_j \rangle=0$) is then computed from $i \hbar G_j$.  Numerical results for nanotubes and graphene sheets with Brenner potential are given in Ref.~\onlinecite{jiang-prb80-205429}.  It is interesting to note that the thermal expansion coefficient in the radial direction of nanotubes and graphene sheets are negative at room temperature or below.  The same method is applied to the study of thermal contraction in silicon nanowires \cite{jiang-nanoscale10}, as well as multi-layered graphene \cite{jiang-arxiv1108.5820}. 

Another simple application of the lowest order perturbation expansion of the nonlinear diagrams is the phonon life-times.  If we work in the normal-mode representation so that in the noninteracting system each vibrational mode is diagonal with the retarded Green's function given by $1/\bigl( (\omega + i \eta)^2 - \Omega_q^2\bigr)$, then the effect of nonlinearity can be interpreted as to shift the vibrational frequencies of each mode $q$ and to give a damping or finite life-time of the mode:
\begin{equation}
G^r_q[\omega] = {1 \over  (\omega + i \eta)^2 - \Omega_q^2 - \Sigma_{n,q}^r[\omega] }.
\end{equation}  
We'll assume that the effect of nonlinearity is small and the retarded Green's function is essentially peaked around $\Omega_q$ with a small shift by a complex number 
$\Delta_q -  i /\tau_q$ where $\tau_q$ is the phonon life-time of mode $q$.  Under such approximation, we can identify the real and imaginary part of the retarded nonlinear self-energy as \cite{maradudin62,yxuprb08}
\begin{eqnarray}
{\rm Re}\, \Sigma^r_{n,q}[\Omega_q] \approx 2 \Omega_q \Delta_q, \\
{\rm Im}\, \Sigma^r_{n,q}[\Omega_q] \approx  - \frac{2 \Omega_q}{\tau_q}.
\end{eqnarray}
It is interesting to note that at the lowest order of approximation from NEGF, the result agrees exactly with the Fermi-Golden rule result.

We can use the kinetic theory formula to compute the thermal conductivity, $\kappa = \sum_q \frac{1}{3} c_q v_q^2 \tau_q$, once the phonon life-time is known, where $c_q$ is heat capacity per unit volume of mode $q$, and $v_q$ is the group velocity of mode $q$.   However, such approaches are not very rigorous, as many assumptions have gone into it.

\section{Nonlinear Systems, Mean-Field Approximation}
Nonlinear problems are the heart of matter and the holy grail of thermal transport.  In principle, NEGF solved the problem formally by giving the nonlinear self-energy $\Sigma_n$ together with the Meir-Wingreen formula for current.  However, a practical calculation with good accuracy is immensely difficult, particularly if the sizes of the systems are large.   As a start, we can use lowest order perturbative expressions for the nonlinear self-energies.  It works to some extent for weak nonlinearity and small sizes.  Our experience seems to indicate that, with just the perturbative terms, it is not possible to produce correct diffusive transport for large sizes \cite{jswprb06,jswpre07}.  The next step is to use self-consistent Born approximation \cite{mingo06,jswpre07,luisier12}, keeping only the lowest order self-energy diagrams (see Fig.~5 in Ref.~\onlinecite{wang08review} for the self-energy diagrams).  Self-consistency means that the ballistic Green's function $G^0$ is replaced by the full Green's function $G$.
Such an approach gives only qualitatively correct results (such as diffusive behavior
for large sizes, i.e., the current decreases with sizes as $1/L$).  One technical difficulty in the self-consistency procedure is that the iterations may not converge, as the Green's functions are oscillatory functions, nonsmooth and ill-behaved.  
Some approximations do not conserve energy exactly, i.e., $I_L + I_R \neq 0$, in steady states \cite{lu07,myohanen09}.
Thus, an accurate, quantitatively correct theory for nonlinear quantum thermal transport in a large parameter region (system sizes, nonlinear strength) is still lacking \cite{lawu11}.   Many calculations are still based on solving the Boltzmann equations \cite{lindsay09} which is semi-classical (because simultaneous position and lattice momentum of phonon distribution are used and certain coherent wave nature is neglected \cite{mlleek12}).

Surprisingly, the mean-field theory under certain restricted condition (quartic nonlinear, small number of degrees of freedom) gives very accurate results in comparison with other methods, in particular, in comparison with the quantum master equation \cite{lfzhang-juzar13}.  The quartic nonlinear model, with the
potential of Eq.~(\ref{eq-Tijkl}),
is a better model to study as it is stable with proper choice of the coefficient $T_{ijkl}$, while the cubic nonlinear term is unstable for sufficiently large displacement and always needs a quartic term to stabilize the system,  although for almost all practical systems, the cubic term should be present.

Our motivation here is to derive reasonably accurate equations for the Green's functions without any ad hoc approximation for the nonlinear self-energy, but rather look at the Green's functions directly.  Since the Green's functions will form a hierarchy, we need to introduce a general $n$-point
Green's function
\begin{equation}
G(1,2,\cdots,n) = -\frac{i}{\hbar} \langle T_c u(1) u(2) \cdots u(n)\rangle,
\end{equation}
where the number denotes the complete set of space index and contour time variables, e.g., 1 means $(j_1, \tau_1)$.  By definition, $G$ is completely symmetric with respect to the permutation of the arguments.   For the quartic potential, Green's functions with an odd number of displacement fields is 0, so we only need to consider $n$ even.  The lowest one is the two-point Green's function.  Applying the equation of motion method, also taking care that we are only interested in the Green's functions involving the center degrees of freedom, we obtain
\begin{eqnarray}
\label{eq-1stbbgky}
\left[ \left( I \frac{\partial^2 \; }{\partial \tau_1^2} + K^C + \Sigma \right) G(\tau_1, \tau_2) \right]_{j_1j_2} = - \delta(\tau_1, \tau_2) \delta_{j_1j_2} \nonumber \\
- \sum_{j_3j_4j_5}T_{j_1j_3j_4j_5} G_{j_3j_4j_5j_2}(\tau_1, \tau_1,\tau_1,\tau_2) , \qquad
\end{eqnarray}
where $I$ is the identity matrix, $K^C$ is the force constant matrix for the center, and $\Sigma$ is the self-energy of the leads.  This equation is exact, and is the first of the BBGKY hierarchy relating $G(1,2)$ to $G(1,2,3,4)$. 
In contour time, the action of matrix $\Sigma$ on $G$ is a convolution, thus
\begin{equation}
\bigl( \Sigma G \bigr) (\tau, \tau') = \int_C  \Sigma(\tau, \tau'') G(\tau'', \tau') d\tau''.
\end{equation}

Equation~(\ref{eq-1stbbgky}) can be ``solved'' or put into an integral form, to give
\begin{eqnarray}
\label{eq-1stbbalt}
G(1,2) &=& G^0(1,2) +  \\ 
&& \int d3\,d4\,d5\,d6\, G^0(1,3) T(3,4,5,6)G(4,5,6,2), \nonumber
\end{eqnarray}
where the contour-time dependent coupling is defined by inserting three contour
$\delta$ functions so that all the times are synchronized, and $G^0$ is the Green's function for the ballistic system (when the nonlinear term is 0).  We can carry on to derive equations for $G(1,2,3,4)$, which will then involve a 6-point Green's function.   At some point, we have to close the equations by certain approximation. As there is no particular good reason to prefer one approximation over the other, such an approach is overly complicated and seems at a loss.    Perhaps the simplest one among all is to stop as early as possible, thus we consider
\begin{eqnarray}
\left(-\frac{i}{\hbar}\right)G(1,2,3,4) &\approx& G(1,2)G(3,4) +  \\ 
&& G(1,3)G(2,4) + G(1,4)G(2,3). \nonumber
\end{eqnarray}
This equation would be exact if Wick's theorem is valid.  This approximation is amount to the assumption that high-order correlations (4-th order) are small, thus taking only 2-point correlations should be already a good approximation.  We like to point out that this is not a weak nonlinear approximation, as it is not obtained by truncating a perturbation series. The validity of such an approximation can only be tested numerically.  Putting this approximation for the 4-point Green's function back into Eq.~(\ref{eq-1stbbgky}) or (\ref{eq-1stbbalt}) we see that the result is equivalent to having a self-consistent nonlinear self-energy, taking only the lowest order diagram
\begin{equation}
\Sigma_n(\tau,\tau')_{jj'} = 3 i \hbar\, \delta(\tau, \tau') \sum_{kl} T_{jj'kl}G_{kl}(\tau,\tau).
\end{equation}
We note that this nonlinear self-energy is real, thus only shifting the frequencies of the modes, and hence it cannot give a finite life-time for the phonons, unable to describe diffusive transport.  This NEGF version of ``effective phonon'' theory is closely related to the effective phonon theory of He et al., where the temperature-dependent force constants are derived based on Feynman-Jensen inequality \cite{dhe-pre08}.
We'll call this version of mean-field theory as the self-consistent mean-field (SCMF).

\begin{figure}
\includegraphics[width=\columnwidth]{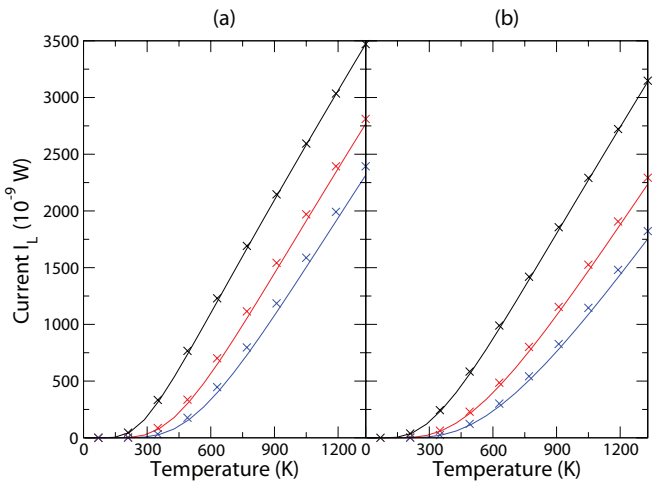}%
\caption{\label{fig:scmf}(Color online) Comparison between SCMF (solid lines) and master equation (crosses) for the one (left) and two (right) particle quartic nonlinear model.  For the one particle case $\Omega^2 = 60.321$ meV/(\AA$^2$u) and $T_{1111} = 0.241$ (black curve), 1.2 (red curve), 2.4 (blue curve) [eV/(\AA$^4$u$^2$)].  In case of two particles, $K_{11}=K_{22}=60.321$, $K_{12}=K_{21} = - 30.165$ [meV/(\AA$^2$u)]; $T_{1111} = T_{2222} = 0.483$, $T_{\{1,1,1,2\}} = -T_{\{1,1,2,2\}} =- 0.241$ (black curve); $T_{1111} = T_{2222} = 2.4$, $T_{\{1,1,1,2\}} = -T_{\{1,1,2,2\}} =- 1.2$ (red curve); $T_{1111} = T_{2222} = 4.8$, $T_{\{1,1,1,2\}} = -T_{\{1,1,2,2\}} =- 2.4$ (blue curve)  [eV/(\AA$^4$u$^2$)] and the curly-brackets in subscripts indicate all possible permutations of the indices.  The retarded self-energy of the leads $\Sigma^r_{\alpha}[\omega] = \frac{1}{\pi} {\rm P} \int_{-\infty}^{+\infty} J_{\alpha}(\omega')/(\omega - \omega')d\omega' - i J_{\alpha}(\omega)$, with $J_{\alpha}(\omega) = \epsilon^2 \omega/(1 + \omega^2/\omega_D^2)$, $\alpha=L,R$, $\epsilon = 6.0321$ meV/(\AA$^2$u), $\hbar \omega_D = 10$ eV, $T_L = 1.25 T$ and $T_R = 0.75T$, which corresponds to the Lorentz-Drude model of heat baths.}
\end{figure}

The current can still be calculated using Caroli/Landauer formula with $\Sigma_n^r$ incorporated in $G^r$. 
Figure~\ref{fig:scmf} shows the comparison between SCMF (solid lines) and master equation approach (crosses) in the weak system-bath coupling regime \cite{juzar-thesis}.  Since the master equation approach becomes computationally very demanding for the number of particles $\ge 3$, we restrict our comparison to one and two particle systems as shown in Fig.~\ref{fig:scmf} (a) and (b), respectively. As the master equation formulation makes no assumptions for the strength of the anharmonicity, it should be considered as a numerically exact result, bearing in mind that the system-bath coupling is weak.  Surprisingly, the SCMF approach matches the master equation formulation for very strong values of anharmonicity indicating that the self-consistency procedure is probably one of the key ingredients for treating strongly anharmonic systems within the NEGF framework.

\section{Reduced density matrix for ballistic systems}
For ballistic systems, we can calculate the $n$-point Green's functions (for the center degrees of freedom) for any values of $n$.  Clearly, the set of all Green's functions completely characterizes the steady state of a nonequilibrium system.  In fact, only two-point Green's function is needed as the higher order ones reduce to the two-point one by the validity of Wick's theorem in a ballistic system.   Alternatively, the reduced density matrix also completely characterizes a nonequilibrium steady state.  The reduce density matrix, obtained by tracing over the bath degrees of freedom, is a better (local) quantity to define steady state.  For the full density matrix, some sort of limit of bath degrees going to infinity and time going to infinity has to be taken in order to reach steady state, but such limits may not be well-defined.

In this section, we present the method of Dhar, Saito, and H\"anggi \cite{dharsh12} who gave a procedure to compute the reduced density matrix in a nonequilibrium steady state.   The starting point is an ansatz that the reduced density matrix, although unknown, must be quadratic in the basic dynamic variables $u_C$ and $p_C$ of the center.  For notational simplicity, we'll drop the subscript $C$ in the following.  We start by defining a vector $\varphi = (u,p)^T$, then we can write the reduced density matrix of the center as
\begin{equation}
\rho \propto \exp( -\varphi^T\! A\, \varphi),
\end{equation}
where $A$ is a matrix with twice the size of the degrees of freedom of the center, and the proportionality constant can be fixed by normalization, ${\rm Tr}( \rho )= 1$.  We note that the equilibrium statistical mechanics Gibbs distribution, $\exp(-\beta H_C)$, is also of this form.  The nonequilibrium distribution introduces mixing terms between $u$ and $p$.  We determine $A$ by matching the Green's functions.   It is not necessary to use the complete time-displaced Green's functions.  It is sufficient just to use the static ones, i.e., the Green's functions at equal time, or the covariance matrix
\begin{equation}
C = \langle \varphi \varphi^T \rangle = 
\left( \begin{array}{cc}
              \langle u\,u^T \rangle & \langle u\,p^T \rangle \\
              \langle p\,u^T \rangle & \langle p\,p^T \rangle  
              \end{array}
     \right),
\end{equation} 
where the angular brackets mean trace with respect to the reduced density matrix
$\rho$.   The $uu$ correlation can be computed with the greater or lesser Green's functions at time 0 or using the integral of Green's function in frequency domain, e.g.,
\begin{equation}
\langle u u^T \rangle = i\hbar G^<(0) = i\hbar \int_{-\infty}^{+\infty} \!\!\frac{d\omega}{2\pi}\,
G^<[\omega].
\end{equation}
Using the Keldysh equation $G^< = G^r \Sigma^< G^a$,  $\Sigma^< = \Sigma_L^< + \Sigma_R^<$, and $\Sigma_\alpha^< 
= - i f_\alpha \Gamma_\alpha$, $\alpha = L, R$ for the two-lead situation, the above expression can be evaluated.  The terms involving momenta (velocities in our convention of unit mass) can also be computed by noting  $ \partial \langle u(t) u(t')^T \rangle /\partial 
t = \langle p(t) u(t') \rangle= i \hbar \partial G^>(t,t') /\partial t$.  Thus, in frequency domain, each derivative introduces an extra $\pm i \omega$ factor. 

If $\varphi$ were just ordinary vector of numbers, and the distribution is a general Gaussian, it is easy to see that $A = \frac{1}{2} C^{-1}$.  This is not true when $\varphi$ are operators.  An important step in Ref.~\onlinecite{dharsh12} is  the introduction of a  linear symplectic (or canonical) transform, $\phi' = S \varphi$, which preserves the commutation relations, and satisfies
\begin{eqnarray}
S J S^T = J  = \left( \begin{array}{cc}
              0 & I \\
              -I & 0  
              \end{array}
     \right), \qquad\qquad\qquad &\\
S \frac{1}{2}\left( C + C^T\right) S^T = {\rm Diag}(d_1, \cdots, d_N, d_1, \cdots, d_N), \qquad & \\
A = S^T {\rm Diag}(a_1, \cdots, a_N, a_1, \cdots a_N) S, \qquad \qquad& 
\end{eqnarray}
where $I$ is the identity matrix. $S$ is chosen such that the symmetrized $C$ matrix is diagonalized with diagonal elements $d_s$ repeated twice.  Simultaneously, $S$ also diagonalizes $A$ with diagonal elements $a_s$.  A numerical procedure to do this is given in appendix of Ref.~\onlinecite{dharsh12}.  Since in variable $\phi'$ the system is diagonal, the problem becomes equivalent to the problem of a set of decoupled harmonic oscillators each at a certain effective temperature.  This gives the relation 
\begin{equation}
d_s = \frac{\hbar}{2} \coth(\hbar a_s).
\end{equation}
The exact expression for the reduced density matrix offers a good way to compare with the quantum master equation approach, which will be discussed in the next section.

\section{Quantum Master equations}
In this last section before conclusion, we take a look at the quantum master equation approach \cite{breuerpetruccione02} to thermal transport \cite{saitoepl03,segalprl05,segalprb06} from the point of view of NEGF.  Although NEGF is a complete theory for answering the questions of thermal currents and other observables, it is still very difficult to handle nonlinear systems in general.  On the other hand, the quantum master equation approach handles nonlinearity with great ease: any finite degree center is treated the same way by expanding in the eigenstates of the center.  But the price we have to pay is that we cannot treat the couplings between the baths and center exactly.  However, if we can systematically improve  the weak-coupling approximation, then the master equation approach offers a great advantage.  

Just like NEGF, the master equations have a long history \cite{pauli28,caldeiraleggett83}.  The most commonly used ones keep the system-bath coupling accurate in second order \cite{redfield57,lindblad76}. Such equations can give steady state solutions for the density matrix accurate to the lowest, zeroth order only \cite{mori08,fleming11}.  Some progress is made recently \cite{thingna-jcp12} in extending the accuracy of the density matrix to second order by a novel analytic continuation without actually solving more complicated fourth order master equation \cite{laird91,sjang01}. Formally exact quantum master equation exists either in a time-nonlocal form \cite{nakajima58,zwanzig60} or time-local form \cite{shibata77,nanJCP09}.  Here we give a transparent derivation of the higher order time-local master equation as well as energy current, using the contour order as a connection point to NEGF. 

We'll restrict the scope to nonequilibrium steady state only, although a generalization to time-dependent dynamics is straightforward.  There are two quantities of interests --- one is the reduced density matrix $\rho$, and the other is the current $I_L$.  Both of them can be treated in a similar way.  The standard approach for obtaining the steady state is to evolve from the remote past product initial state adiabatically (or even suddenly as we did in the earlier part of this review).  In NEGF, there is no problem with this approach since the couplings between the leads and the center system are handled exactly.   However, in the master equation approach (specifically, when the couplings themselves are treated perturbatively), this adiabatic switch-on results in divergences for both the reduced density matrix and the current beyond the lowest order, generically for any initial product state $\rho_0 \rho_B$ where $\rho_0$ denotes the density matrix of the center and $\rho_B=\rho_L \rho_R$ for the equilibrium baths.   Unless the initial state is carefully chosen, a steady state cannot be reached.  

The way to overcome this divergence is to impose a steady state condition for the initial state, by the requirement that the rate of change of the reduced density matrix $\rho$ should be 0.  With this, the initial state $\rho_0$ is determined from a condition, rather than an initial input that takes any arbitrary value.  Since $\rho_0$ needs to be determined, and formally we can create a unique, invertible map $\rho_0$ to $\rho$, the problem is equivalent to determining an equation for $\rho$, which is commonly known as the master equation. 

The one-to-one map $\rho_0 \leftrightarrow \rho$ exists only before the adiabatic limit ($\epsilon \to 0^+$) is taken. Then our recipes are the followings: (1) give a formal Dyson expansion result for the physical quantity $\langle O \rangle$, in terms of $\rho_0$, where $O$ can be the Hubbard operator $X_{mn} = |m\rangle \langle n|$ (where $|n\rangle$ is the $n$-th eigenstate of the isolated center) or the current operator $I_L = p_L^T V^{LC} u_C$; (2) Using (1) to obtain both $\rho$ and $d \rho/dt$, inverting the relation from $\rho_0$ to $\rho$ and substituting it back into the original physical quantities, we simultaneously obtain  the equation for the current as well as the master equation to any desired order of accuracy in terms of the couplings.

Here we introduce some notations: the total Hamiltonian is $H=h + V$ where $h = H_C + H_L + H_R$ is the decoupled system and $V=u_L^T V^{LC}u_C + u_R^T V^{RC}u_C$ is the bath-system coupling potential.  Working in the interaction picture with respect to $h$, and setting the synchronization time among different pictures to $0$, we have, for any observable
\begin{eqnarray}
\langle O_H(t) \rangle &=& {\rm Tr}\bigl[ \rho_0 \rho_B S(-\infty, t) O(t) S(t,-\infty) \bigr] \nonumber \\
\label{eq:Oht}
&=& {\rm Tr}\left[ \rho_0 \rho_B T_c \left\{  O(t) e^{\lambda \int_C V(\tau)d\tau} \right\} \right],
\end{eqnarray}
where $O(t) = e^{ith/\hbar} O e^{-ith/\hbar}$ is the observable in the interaction picture, while $O_H(t)$ is the same observable in the Heisenberg picture.
The contour $C$ now runs from $-\infty$ to time of interest $t$.  In $V(t)$ we have included implicitly an adiabatic switch-on parameter $e^{\epsilon t}$ in addition to the usual interaction picture form of $V$.  The parameter $\lambda =(-i/\hbar)$ serves as a small expansion parameter in a Dyson expansion (we could also absorb a small parameter of the coupling from $V$ into $\lambda$).   We have assumed that $[\rho_0,H_C]=0$ for the validity of Eq.~(\ref{eq:Oht}), but this is not a fundamental limitation.  We can always use $\rho_0' = e^{iht_0/\hbar} \rho_0 e^{-iht_0/\hbar}$ with a finite $t_0$ instead of $t_0 \to -\infty$.  Since $\rho_0$ or $\rho_0'$ is eliminated in the end, the results below are independent of this assumption.
We also note that the rate of change of $O$ at time $t$ is given by 
\begin{eqnarray}
\frac{d\;}{dt} \langle O_H(t) \rangle 
&=& {\rm Tr}\Bigg[ \rho_0 \rho_B T_c \Big\{  \bigl(\dot{O}(t) + \nonumber \\
\label{eq-dO}
&& \lambda\bigl[O(t),V(t)\bigr]\bigr) e^{\lambda \int_C V(\tau)d\tau} \Big\} \Bigg],
\end{eqnarray}
where $\bigl[O(t), V(t)\bigr]$ is the commutator of the two operators.  We'll use the symbol $X$ to denote a matrix with matrix element $X_{nm} = |n \rangle \langle m|$.  Then the reduced density matrix at time $t=0$ can be computed as $\rho = \langle X_H^T(0) \rangle$ and its derivative can also be similarly computed.  
Performing the power series expansion for the exponential, and noting that an odd number of bath operators $u_L$ or $u_R$ gives 0, we obtain
\begin{equation}
\rho = \langle X^T\rangle + \frac{\lambda^2}{2!}
\langle X^T V^2 \rangle + \frac{\lambda^4}{4!}\langle X^T V^4\rangle + O(\lambda^6),
\end{equation}
where we have introduced a short-hand notation of the angular brackets to mean 
${\rm Tr}\bigl[ \rho_0 \rho_B T_c \int_C \cdots \bigr]$ and where the number of contour integrals depends on the number $n$ in $V^n = V(\tau_1) V(\tau_2) \cdots V(\tau_n)$.  As the first term is explicitly $\rho_0 = \langle X^T \rangle$, it is possible to invert this equation, to express $\rho_0$ in terms of the final $\rho$, giving up to 6-th order, as
\begin{eqnarray}
\rho_0 &=& \rho - \frac{\lambda^2}{2} \langle X^T V^2\rangle_\rho 
-\frac{\lambda^4}{4!}\langle X^T V^4 \rangle_\rho + \nonumber \\
&& \frac{\lambda^4}{2!\,2!} \bigl\langle \langle X^T V^2 \rangle_\rho X^T V^2 \bigr\rangle - \frac{\lambda^6}{6!} \langle X^T V^6\rangle_\rho  + \nonumber \\
&&\frac{\lambda^6}{2!\,4!} \bigl\langle \langle X^T V^2 \rangle_\rho X^T V^4 \bigr\rangle +  
\frac{\lambda^6}{4!\,2!} \bigl\langle \langle X^T V^4 \rangle_\rho X^T V^2 \bigr\rangle \nonumber \\ 
\label{eq-rho0}
&&-\frac{\lambda^6}{(2!)^3} \bigl\langle \bigl\langle \langle X^T V^2 \rangle_\rho X^T V^2 \bigr\rangle X^T V^2 \bigr\rangle  + O(\lambda^8),  
\end{eqnarray}
where the meaning of the angular brackets are changed again.  The  brackets at the innermost level with a subscript $\rho$ is the same as before expect that $\rho_0$ is replaced by $\rho$. The outer slightly larger angular brackets means a trace over the density matrix of the bath, $\rho_B$, as well over the system with a matrix produced by $\langle \cdots \rangle$ inside it.  In addition, we still have the contour integrals.  For example, let $\rho^{(2)} = \langle X^T V^2\rangle_{\rho}$, i.e., $\rho^{(2)}_{nm} = {\rm Tr}\bigl[\rho \rho_B |m\rangle \langle n|  \int d\tau_1 \int d\tau_2 T_c V(\tau_1) V(\tau_2)  \bigr]$, then the second $\lambda^4$ term is  $\bigl\langle \langle X^T V^2 \rangle_\rho X^T V^2 \bigr\rangle_{nm} = 
{\rm Tr}\bigl[\rho^{(2)} \rho_B |m\rangle \langle n|  \int\! d\tau_1 \int\! d\tau_2 T_c V(\tau_1) V(\tau_2)  \bigr]$, which is, implicitly, a linear function of $\rho$.

The derivative of the density matrix, $d\rho /dt$, can be similarly expanded using Eq.~(\ref{eq-dO}), and when $\rho_0$ is substituted with Eq.~(\ref{eq-rho0}), we formally obtain the time-local master equation \cite{laird91,sjang01}, up to 6-order, as
 \begin{eqnarray}
{ d\rho \over dt} &=&  -i \Delta \cdot \rho + 
\lambda^2 \langle [X^T,V]V \rangle_{\rho} + 
\frac{\lambda^4}{3!} \langle [X^T,V]V^3 \rangle_{\rho} \nonumber \\
&&-\frac{\lambda^4}{2!} \bigl\langle \langle X^TV^2\rangle_{\rho} [X^T,V] V \bigr\rangle + 
\frac{\lambda^6}{5!} \langle [X^T,V]V^5 \rangle_{\rho}
\nonumber \\
&&-\frac{\lambda^6}{4!} \bigl\langle \langle X^TV^4\rangle_{\rho} [X^T,V] V \bigr\rangle + \nonumber \\
&&+\frac{\lambda^6}{2!\,2!} \bigl\langle \bigl\langle \langle X^TV^2\rangle_{\rho} X^T V^2 \bigr\rangle [X^T,V] V \bigr\rangle + \nonumber \\
&& 
-\frac{\lambda^6}{2!\,3!} \bigl\langle \langle X^TV^2\rangle_{\rho} [X^T,V] V^3 \bigr\rangle + O(\lambda^8) = 0.
\end{eqnarray}
The meaning of the first term is $-i \frac{E_n-E_m}{\hbar} \rho_{nm}$, if we write out explicitly in matrix element form, where $E_n$ is the eigen-energy of the $n$-th state of the isolated, arbitrary nonlinear center of $H_C$.
The meaning of the two types of angular brackets remains the same.  The time arguments for $[X^T,V]$ are at $t=0$, while all the other $V$s have dummy contour time argument $\tau_i$ and need to be integrated out.  The $\rho$ dependence is
in the angular brackets $\langle \cdots \rangle_{\rho} = {\rm Tr}[ \rho \rho_B T_c \int_C d\tau \cdots]$.  After performing the trace and contour integrals, we obtain
explicitly the equation for $\rho$. 
If we truncate the series to second order in $\lambda$, we get the standard Redfield quantum master equation \cite{redfield57}. 

The current can be treated in a similar fashion.  In fact, the mathematical structure of the current is the same as that of master equation, except that we just need to replace the commutator, $[X^T,V]$, by $\!\!\dot{\,\;V} = p_L^T V^{LC}u_C$ where the left-sided dot on $V$ indicates that the time derivative is performed only for the left lead, then we can write, for the quantity $I = \lambda I_L$, as
\begin{eqnarray}
\langle I_H \rangle &=& \lambda \langle \!\!\dot{\,\;V} \rangle =
\lambda \langle \!\!\dot{\,\;V} e^{\lambda \int_C \!\!V(\tau) d\tau} \rangle \\
&=& \lambda^2 \langle \!\!\dot{\,\;V}V \rangle + 
\frac{\lambda^4}{3!} \langle \!\!\dot{\,\;V}V^3 \rangle + 
\frac{\lambda^6}{5!} \langle \!\!\dot{\,\;V}V^5 \rangle + \cdots \nonumber \\
&=& \lambda^2 \langle \!\!\dot{\,\;V}V \rangle_{\rho} + 
\frac{\lambda^4}{3!} \langle \!\!\dot{\,\;V}V^3 \rangle_{\rho}
-\frac{\lambda^4}{2!} \bigl\langle \langle X^TV^2\rangle_{\rho} \!\!\dot{\,\;V} V \bigr\rangle + O(\lambda^6). \nonumber
\end{eqnarray}
The last line is due to Eq.~(\ref{eq-rho0}) where $\rho_0$ is written in terms of $\rho$. 
A divergence appearing in the second term gets cancelled explicitly by the third term. This approach solved the divergence problem which has been puzzling us for a while.  The above result is a generalization of the second order result (the first term) in Ref.~\onlinecite{juzar-prb12}, see also Ref.~\onlinecite{wupre09}.

\begin{figure}
\includegraphics[width=\columnwidth]{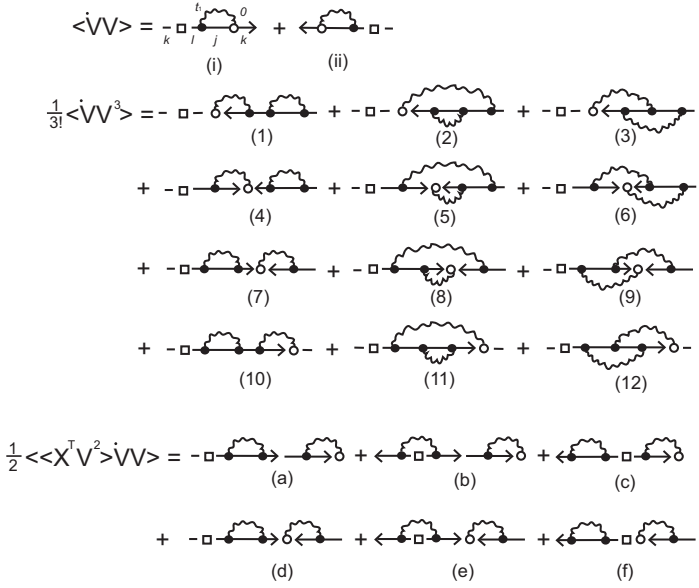}%
\caption{\label{fig:current}Diagrams representing the terms in the current. The graphs (1), (4), (5), (7), (8), (10) and (a)-(f) have divergent terms of the form $\propto 1/\epsilon$. The Feynman rules are discussed in the text.}
\end{figure}

It is possible to represent various terms for the expression of current in terms of diagrams after unraveling the contour time into normal time with time order or anti-time order.  The diagrams are shown for the various terms in Fig.~\ref{fig:current}.   The Feynman rules for the diagrams are as follows: 1) each dot is associated with a time $t_j$ and the system operator; each line segment between the dots has a system state label; the matrix element of the operator is
$\langle k | u_C | l\rangle e^{i \Delta_{kl} t_j + \epsilon t_j}$, where
we define $\Delta_{kl} = (E_k - E_l)/\hbar$.  2) The dots are connected by the phonon lines, representing the contour function $C(\tau_1, \tau_2) = i\hbar \Sigma(\tau_1, \tau_2)$, $\tau_j = t_j^\sigma$, in all possible ways.  3) The open dot has a fixed time of 0, and associated $\!\!\dot{\,\;C}$ has a time derivative. 4) The reduced density matrix $\rho$ is represented as a square box.  5) The direction of the arrow represents ordering; right pointing arrows are for anti-time order (lower branch), and left pointing arrows time order (upper branch).  The horizontal line represents the trace over the system states.  With these rules, e.g., the first two diagrams, (i), (ii), and diagram (3) are (assuming center has only one degree of freedom and $S_{kl} =\langle k | u_C | l\rangle$, for multiple degrees of freedom, $C$ becomes a matrix and $S$ received a vector index)
\begin{eqnarray}
({\rm i}) &=& \sum_{klj} \int_{0}^{-\infty}\!\!\!\!\!dt_1\, \rho_{kl}S_{lj}S_{jk} e^{i \Delta_{lj}t_1 + \epsilon t_1} \!\!\dot{\,\;C}(0,t_1^-), \\
({\rm ii}) &=& \sum_{klj} \int_{-\infty}^{0}\!\!\!\!\!dt_1\, \rho_{kl}S_{lj}S_{jk} e^{i \Delta_{jk}t_1 + \epsilon t_1} \!\!\dot{\,\;C}(0,t_1^+), \\
({\rm 3}) &=& \sum_{klpqj} \rho_{kl} S_{lp}S_{pq}S_{qj}S_{jk} \int_{-\infty}^0\!\!\!\!dt_1
\int_{-\infty}^{t_1}\!\!\!\!dt_2 \int_{-\infty}^{t_2}\!\!\!\! dt_3  \\
&& e^{i\Delta_{pq}t_1 + i\Delta_{qj}t_2 + i\Delta_{jk}t_3 + \epsilon(t_1 + t_2 +t_3)} \!\!\dot{\,\;C}(0,t_2^+) C(t_1^+, t_2^+). \nonumber
\end{eqnarray}
We intend to give a full account of this method with numerical application elsewhere \cite{juzar-wang-unpub13}.

\section{Conclusion}

In this review, we gave simple examples of Green's functions for harmonic oscillator as a starting point.  We have tried to emphasize the contour ordered functions as the basic language of NEGF.   The functions are defined on a finite segment from the initial time to the current time of interest.  This formulation makes the theory work equally well for steady state and transient time development.   A number of applications explore this feature, noticeably the transient problem of full counting statistics.  The standard topics of NEGF, the equation of motion method and Feynman-diagrammatic expansion, were briefly discussed.  Some recent developments are reviewed, such as the new formulas for coupled left-right leads.  In the treatment of nonlinear systems, we draw attention to the self-consistent mean field theory, which was shown to give surprisingly accurate result for the current for small systems.   We attempted to make connections between NEGF and master equation.  The master equation approach focuses on the reduced density matrix.  An exact expression for the density matrix for ballistic system is reviewed.  The last section is a bit off the main line of this review.  There, we gave a very transparent derivation of the higher order (time-local) quantum master equation and a suggestion on how higher order current can be computed.  This last part is new, to our knowledge.

\section*{Acknowledgments}
J.-S. W thanks Hong Guo, from whom the Caroli formula was first exposed, and  Lin Yi where the book of Haug and Jauho and work of Meir and Wingreen were made known to the author.  He also thanks Jian Wang, Jingtao L\"u, Jin-Wu Jiang, Eduardo Cuansing, Lifa Zhang, Jinghua Lan, and Baowen Li for many collaborations.

\vfill
\bibliography{thermal,NEGF-2013}

\end{document}